\documentclass[draft]{agujournal2019}
\usepackage{url} 
\usepackage{lineno}
\usepackage{amsmath}
\usepackage{soul}
\draftfalse

\journalname{Geophysical Research Letters}

\begin{document}

\title{Samudra: An AI Global Ocean Emulator for Climate} 


%
%

\authors{Surya Dheeshjith\affil{1}, Adam Subel\affil{1}, Alistair Adcroft\affil{2}, Julius Busecke\affil{4}, Carlos Fernandez-Granda\affil{1,3}, Shubham Gupta\affil{1}, Laure Zanna\affil{1,3}}

\affiliation{1}{Courant Institute of Mathematical Sciences, New York University}
\affiliation{2}{Program in Atmospheric and Oceanic Sciences, Princeton University}
\affiliation{3}{Center for Data Science, New York University}
\affiliation{4}{Lamont Doherty Earth Observatory, Columbia University}

\correspondingauthor{Surya Dheeshjith}{sd5313@nyu.edu}


\begin{keypoints}
\item We develop a global, 3D, ocean autoregressive machine learning emulator for climate studies.
\item The emulator, based on a UNet architecture, is stable for centuries, producing accurate climatologies and variability of ocean variables.
\item The emulator training is robust to changes in seeds and initial conditions in the data. 
\end{keypoints}

\begin{abstract}
AI emulators for forecasting have emerged as powerful tools that can outperform conventional numerical predictions. 
The next frontier is to build emulators for long climate simulations with skill across a range of spatiotemporal scales, a particularly important goal for the ocean. 
Our work builds a skillful global emulator of the ocean component of a state-of-the-art climate model.
We emulate key ocean variables, sea surface height, horizontal velocities, temperature, and salinity, across their full depth. 
We use a modified ConvNeXt UNet architecture trained on multi-depth levels of ocean data.
We show that the ocean emulator -- \textit{Samudra} -- which exhibits no drift relative to the truth, can reproduce the depth structure of ocean variables and their interannual variability. 
Samudra is stable for centuries and 150 times faster than the original ocean model. 
Samudra struggles to capture the correct magnitude of the forcing trends and simultaneously remain stable, requiring further work.
\end{abstract}

\section*{Plain Language Summary}
AI tools are extremely effective in making fast and accurate predictions on weather to seasonal timescales. Capturing decadal to centennial changes, which arise from ocean dynamics, remains an outstanding challenge. We built an advanced AI model called   ``Samudra" to simulate global ocean behavior.  Samudra is trained on simulated data from a state-of-the-art ocean climate model and predicts key ocean features such as sea surface height, currents, temperature, and salinity throughout the ocean's depth.  Samudra can accurately recreate patterns in ocean variables, including year-to-year changes. It is stable over centuries and is 150 times faster than traditional ocean models.  However, Samudra still faces challenges in balancing stability with accurately predicting the effects of external factors (like climate trends), and further improvements are needed to address this limitation.


%
%

\section{Introduction}

The recent success of emulators for components of the climate system, primarily the atmosphere, continues to produce remarkable outcomes,  achieving state-of-the-art performance for weather prediction tasks \cite{kochkov2024neural,bi2023accurate,price2023gencast} and promising results reproducing climate models over decadal \cite{cachay2024probabilistic} to multi-decadal timescales \cite{watt2023ace}.  

Existing work on ocean emulation has mainly been limited to the surface and upper ocean, or to steady forcing.
Several works focusing on surface ocean variables show results for timescales of years to a decade \cite{subel2024building,dheeshjith2024transfer,gray2024long}. 
Emulators that include subsurface information have focused on the weekly to decadal timescales and at most the upper 1000$\mathrm{m}$  \cite{xiong2023ai,guo2024orca,holmberg2024regional,patel2024exploring, arcomano2023hybrid}. \citeA{bire2023ocean} explored longer timescales within a simplified ocean model with idealized steady forcing.
Finally, a seasonal coupled atmosphere-ocean emulator has shown promising results, considering the upper 300$\mathrm{m}$ of the ocean \cite{wang2024coupled}. These ocean and atmosphere emulators have been used for seasonal forecasts based on reanalysis data, and to build surrogates of numerical models.

Emulators of traditional numerical climate models leverage the computational efficiency of machine learning approaches to reduce the often prohibitive computational cost of running a large number of simulations on the original (usually CPU-based) climate model. 
One of the main benefits of emulators is the ability to run large ensembles.
Such ensembles can be used to probe the likelihood of extreme events, explore the climate response to a range of forcing scenarios (e.g., greenhouse gases), and facilitate the development of numerical models by reducing the number of perturbed parameter experiments typically used for calibration \cite{maher2021large,mahesh2024huge}.
Emulators can also accelerate the spin-up integration of numerical models or replace full model components in a coupled setting \cite{khatiwala2024efficient}. 
Finally, emulators can help with data assimilation, replacing an expensive numerical model with a fast surrogate to generate affordable ensembles or an approximate adjoint, maintaining accuracy with reduced cost \cite{manshausen2024generative}.

Our goal here is to reproduce the full-depth ocean state for four 3D and one 2D prognostic variables, using a time-dependent realistic atmospheric forcing as input, extending the work of \citeA{subel2024building,dheeshjith2024transfer}.
At rollout lengths of nearly a decade, our emulator shows considerable skill across several key diagnostics (mean and variance) when compared to the parent numerical model output, which is our ground truth.
In particular, both the temperature structure as a function of depth and the El Niño-Southern Oscillation (ENSO) variability are well reproduced by the emulator.

Simultaneously capturing variables with vastly different timescales, such as velocity (which can contain fast fluctuations) and salinity (which typically fluctuates more slowly), is an outstanding issue for long integrations (already encountered by \citeA{subel2024building}).
To alleviate this problem, we introduce an additional emulator by focusing on the thermodynamic variables (i.e. potential temperature and salinity only). 
This additional emulator captures the slowly varying changes in potential temperature and salinity on timescales of decades to centuries.  

We show that our emulator can retain skill and remain stable for centuries for experiments equivalent to both control and climate-change simulations. 
However, we also note that this stability is accompanied by a weak response to climate-change forcing.
This work demonstrates (to our knowledge) the first ocean emulator capable of reproducing the full-depth (from the surface down to the ocean floor) ocean temperature structure and its variability, while running for multiple centuries in a realistic configuration with time-dependent forcing. 

The paper is organized as follows.
We discuss the data and all emulator details in Section~\ref{sec:methods}.
We explore the properties of the trained emulator on a test dataset and report several multi-decadal experiments with a range of climate forcing in Section~\ref{sec:results}. 
We present our conclusions in Section~\ref{sec:discussion}.



\section{Methods}
\label{sec:methods}

We built an autoregressive ocean emulator from data generated by a state-of-the-art numerical ocean simulation.
Below, we describe the data, the emulator, the architecture, and the training and evaluation of the emulator.  

\subsection{Data} 

The data was generated by OM4, \cite{adcroft_2019}, an ocean general circulation model that is the ocean component of the state-of-the-art coupled climate model CM4 \cite{held_2019}.
The circulation model was initialized with hydrography from the World Ocean Atlas \cite{woa2013} and forced with atmospheric reanalysis, following the OMIP-2 protocol, with version 1.4 of the JRA reanalysis \cite{tsujino_2020}.
The model was run for 65 years (1958-2022). 

The ocean prognostic variables are potential temperature ($\theta_O$), salinity ($S$), sea surface height ($\operatorname{SSH}$), oceanic zonal ($u$), and meridional ($v$) velocity components.
The circulation model has 75 degrees of freedom in the vertical for each 3D prognostic variable, which we conservatively remap onto 19 fixed-depth levels of variable thickness - [2.5, 10, 22.5, 40, 65, 105, 165, 250, 375, 550, 775, 1050, 1400, 1850, 2400, 3100, 4000, 5000, 6000]$\mathrm{m}$ to reduce the data size.
We also conservatively coarsen the data in time using a 5-day simple average in geopotential coordinates, averaging over the fastest waves resolved by the circulation model (which originally used a 20-minute time-step).

The native horizontal grid for the data has a nominal resolution of $1/4^\circ$, but is curvilinear and has three poles (grid singularities) inland. 
We further post-process by filtering with an $18\times 18$ cell Gaussian kernel using the gcm-filters package \cite{Loose2022}, and then conservatively interpolate onto a $1^\circ \times 1^\circ$ Gaussian grid using the xESMF package \cite{xesmf}.
Values in land are treated as missing, and missing values are imputed with zeros.
Before conservative spatial interpolation, we interpolate the velocities to the center of each cell using the xGCM package \cite{xGCM} and rotate the velocity vectors so that \textit{u} and \textit{v} indicate purely zonal and meridional flow, respectively.

\subsection{Ocean Emulator}

The variables in the ocean emulator are: 
\begin{enumerate}
    \item The ocean state $\boldsymbol{\Phi} = (\theta_O, S, \operatorname{SSH}, u, v)$, which includes all 19 depth levels.
    We denote the subset of thermodynamics variables as  $\boldsymbol{\Phi}_{\text{thermo}} = ( \theta_O, S, \operatorname{SSH})$, as opposed to the dynamic variables $\boldsymbol{\Phi}_{\text{dynamic}} = (u, v)$.
    \item Atmosphere boundary conditions $\boldsymbol{\tau} = (\tau_u, \tau_v, \operatorname{Q}, \operatorname{Q}_{anom})$, which consist of the zonal, $\tau_u$, and meridional, $\tau_v$, surface ocean stress, and net heat flux downward across the ocean surface $\operatorname{Q}$ (below the sea-ice) and its anomalies $\operatorname{Q}_{anom}$. 
    The net heat flux is a sum of the short- and long-wave radiative fluxes, sensible and latent heating, heat content of mass transfer, and heat flux due to frazil formation (see K4 and K5 of \citeA{griffies_omip_2016} for a precise definition of the variable "$\operatorname{hfds}$").
     The heat flux anomalies are calculated by removing the climatological heat flux computed over the 65-year OM4 dataset.
    
\end{enumerate}

Our emulator, $\mathcal{F}$, is built to autoregressively produce multiple future oceanic states given multiple previous oceanic states. Specifically, we use a 2-input - 2-output model configuration.
Mathematically, 
\begin{align}
    \tilde{\boldsymbol{\Phi}}_{t+(n+1)\Delta t}, \tilde{\boldsymbol{\Phi}}_{t+(n+2)\Delta t} = \mathcal{F}(\tilde{\boldsymbol{\Phi}}_{t+(n-1)\Delta t},\tilde{\boldsymbol{\Phi}}_{t+n\Delta t},\boldsymbol{\tau}_{t+n\Delta t})    
\end{align}
where $n$ is a positive integer and $\tilde{\boldsymbol{\Phi}}$ represents the ocean state predicted by the emulator at time $t$.
A depth-varying land mask is used to set land cells in the model output to zero.
We use OM4 ocean states, $\boldsymbol{\Phi}_{t}$ and $\boldsymbol{\Phi}_{t-\Delta t}$, along with the corresponding atmospheric forcing, $\boldsymbol{\tau}_{t}$, to produce the first  predictions.
Subsequent ocean states are recursively produced by using previously generated ocean states as input.
We illustrate the rollout process of the emulator in Figure \ref{fig:fig1}a).
The use of multiple input states provides additional context to the emulator, similar to the use of model time tendencies in PDE-based numerical integrations. 
In all of our experiments, $\Delta t = 5~\mathrm{days}$.

\subsection{Architecture}

The emulator is based on the ConvNeXt UNet architecture from \cite{dheeshjith2024transfer}, where the core blocks of a UNet \cite{ronneberger2015u}  are inspired by ConvNeXt blocks \cite{liu2022convnet} adapted from \cite{karlbauer2023advancing}.
The UNet implements downsampling based on average pooling and upsampling based on bilinear interpolation, which enables it to learn features at multiple scales.
Each ConvNext block includes GeLU activations, increased dilation rates, and inverted channel bottlenecks.
We did not use inverted channel depths and replaced the large $7 \times 7$ kernels with $3 \times 3$ kernels.
We use batch normalization instead of layer normalization, as it yielded better skill.
The encoder and decoder consist of four ConvNeXt blocks, each with channel widths [200, 250, 300, 400].
The dilation rates used for both the encoder and decoder are [1, 2, 4, 8].
Additionally, we include a single ConvNext block (with channel width 400 and dilation 8) in the deepest section of the UNet before upsampling.
The total number of model parameters is 135M.
We apply periodic (or circular) padding in the longitudinal direction and zero padding at the poles as in \cite{dheeshjith2024transfer}.
 
The architecture is modified from \citeA{dheeshjith2024transfer} to process multiple ocean depth levels (as opposed to surface only).
In the surface ocean emulator, which contains only a single depth level,  each channel is associated with a variable. In the multi-depth ocean emulator, each channel is associated with a variable and a depth level.
Our main emulator $\mathcal{F}_{\text{thermo+dynamic}}$ takes as input four 19-level oceanic variables ($\theta_O, S, u, v$), the surface variable $\operatorname{SSH}$ and four atmospheric boundary conditions ($\tau_u, \tau_v,\operatorname{Q}, \operatorname{Q}_{anom}$). It produces five output variables ($\theta_O, S, \operatorname{SSH}, u, v$).
As discussed above, we use a 2-input 2-output model configuration and thus, there are $(4 \times 19 + 1) \times 2 + 4 = 158$ input and $(4 \times 19 + 1) \times 2 = 154$ output channels. In addition, we build another emulator $\mathcal{F}_{\text{thermo}}$ that only uses the thermodynamic variables, $\boldsymbol{\Phi}_{\text{thermo}}=( \theta_O, S, \operatorname{SSH})$.


\subsection{Training Details} 

\begin{figure}
\noindent\includegraphics[width=\textwidth]{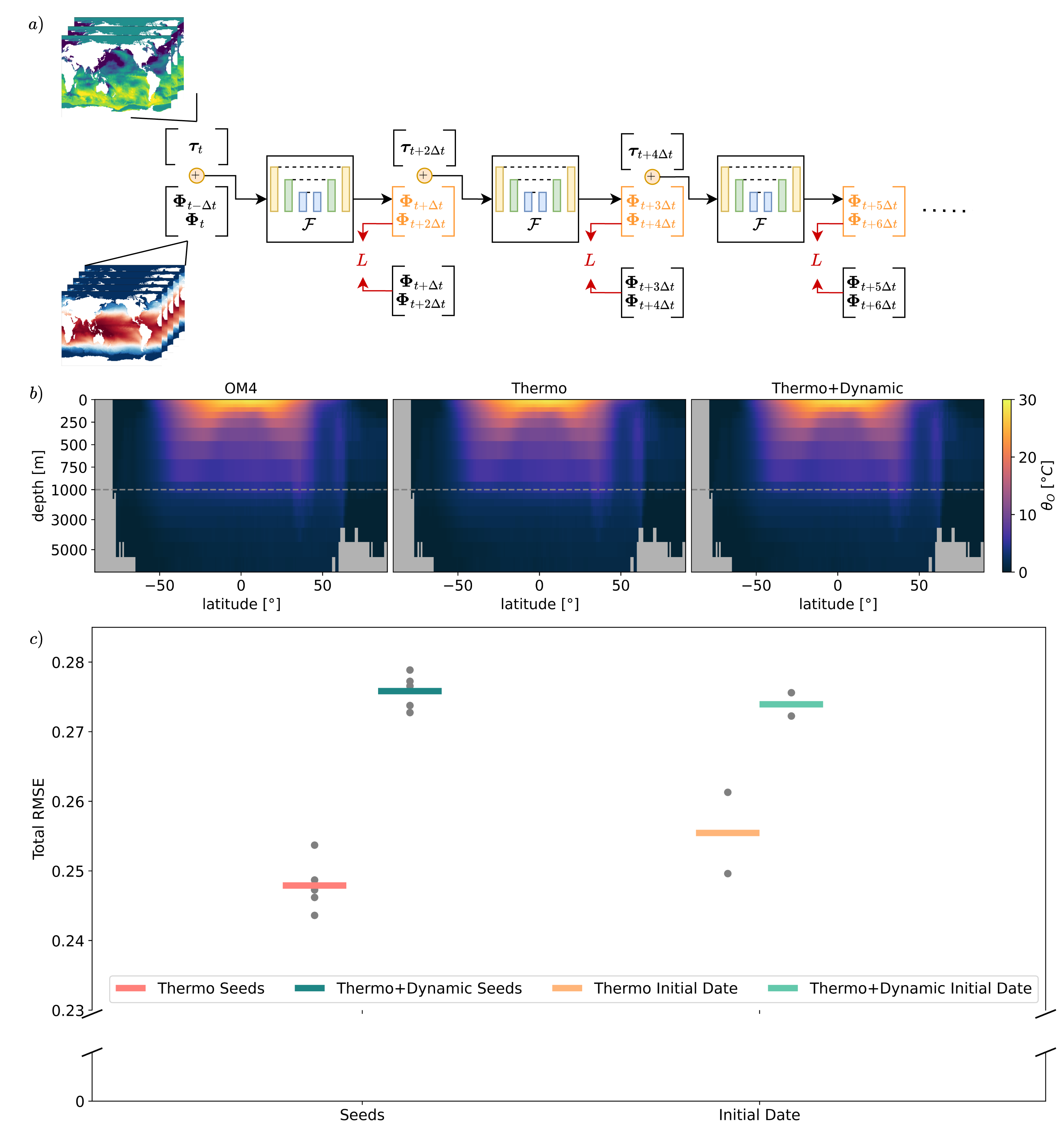}
\caption{a) Schematic of the model training process, illustrating the mapping from input (ocean states and atmospheric forcing) to output (ocean states rolled out over several time steps). Initially, the ground-truth ocean states, $\boldsymbol{\Phi}_{t}$ and $\boldsymbol{\Phi}_{t-\Delta t}$, along with the atmospheric forcing, $\boldsymbol{\tau}_{t}$, are provided as inputs to predict $\boldsymbol{\tilde \Phi}_{t+\Delta t}$ and $\boldsymbol{\tilde \Phi}_{t+2\Delta t}$. Predictions, along with ground-truth atmospheric forcing, are then used as inputs for future steps in the unrolling process. b) Time-averaged potential temperature ($\theta_O$) depth-latitude profiles over the 8-year test set, comparing the ground truth OM4 (left) and predictions from $\mathcal{F}_{\text{thermo}}$ (middle) and $\mathcal{F}_{\text{thermo+dynamic}}$ (right). c) RMSE of 8-year test set predictions for different initial conditions of the emulators, $\mathcal{F}_{\text{thermo}}$ and $\mathcal{F}_{\text{thermo+dynamic}}$. Grey dots represent an RMSE instance of a single rollout, including runs from training on 5 unique model seeds per emulator and 2 additional rollouts initialized at states 6 months apart. Horizontal lines indicate the respective mean RMSE. RMSE is calculated over the common periods of each rollout.}
\label{fig:fig1}
\end{figure}

We illustrate the training of the model in Figure \ref{fig:fig1}a).
We train the emulators using 2900 data samples corresponding to the range 1975-01-03 to 2014-09-20 with the last 50 samples used for validation. Each sample is a 5-day mean of the full ocean state and atmospheric boundary conditions.
 
We ignore data over 1958-1975 due to the excessive model cooling, while it adjusts from the warm initial conditions.
This cooling does not reflect the forcing but rather an interior ocean model adjustment (see \citeA{sane2023parameterizing} and S3).
Note that some regions are still cooling post-1975 in this simulation, which biased some of our testing (see results).

The loss function used for optimization is 
\begin{align}
\mathcal{L}_{t}= \sum_{n=1}^{PN} \frac{1}{C~Y~X} \sum_{j=1}^{C} \sum_{k=1}^{Y} \sum_{l=1}^{X} \left(\boldsymbol{\tilde \Phi}_{t+n\Delta t}^{[j,k,l]} - \boldsymbol{\Phi}_{t+n\Delta t}^{[j,k,l]} \right)^2.
\end{align}
\noindent  $\mathcal{L_{\mathrm{t}}}$ is the total mean square error (MSE) loss function at time step t, where $P$ corresponds to the total number of input/output states used by the model in a single step, $N$ is the total number of recurrent passes, $C$, $Y$ and $X$ are the total number of output channels, height and width, respectively, of a single output state. Here, we set $P = 2$ to obtain a 2-input 2-output model configuration and $N = 4$ steps.

We use the Adam optimizer with a learning rate of $2e-4$, which decays to zero using a Cosine scheduler.
Our emulators are trained using 4 80GB A100 GPUs for 15 and 12 hours for the models $\mathcal{F}_{\text{thermo+dynamic}}$ and $\mathcal{F}_{\text{thermo}}$ respectively, with a total batch size of 16.

\subsection{Evaluation}
\label{sec:Evaluation}

To evaluate the emulators, we take our initial conditions from 2014-09-30 and produce an 8-year rollout using the corresponding atmospheric forcing.
We compare the output from this rollout to held-out OM4 data to evaluate the emulator skill. 
In addition, we produce longer runs to assess the emulator's response, similar to control simulations, with arbitrarily long rollouts.  
The emulator is forced with atmospheric boundary conditions taken from 1990-2000, with a repeat 10-year cycle. This period is chosen specifically because it has a near-zero globally integrated heat flux forcing, which ensures minimal ocean drift. We also performed a 100-year and a 400-year control run (see SI).

We produce predictions using both $\mathcal{F}_{\text{thermo+dynamic}}$ and $\mathcal{F}_{\text{thermo}}$. All evaluations use a single 40GB A100 GPU.
For each year of rollout, $\mathcal{F}_{\text{thermo+dynamic}}$ and $\mathcal{F}_{\text{thermo}}$ take about 90.52$\mathrm{s}$ and 47.2$\mathrm{s}$, respectively.
Thus, for the faster emulator, a century rollout takes approximately 1.3 hours. $\mathcal{F}_{\text{thermo}}$ takes approximately half the time to produce the same number of states in the rollout compared to $\mathcal{F}_{\text{thermo+dynamic}}$.

\section{Results}
\label{sec:results}

\subsection{Full-depth Global Ocean Emulator}

We begin by evaluating the emulators $\mathcal{F}_{\text{thermo+dynamic}}$ and $\mathcal{F}_{\text{thermo}}$ against the ground truth to establish a baseline skill.
Capturing the full-depth climatological profiles of potential temperature and salinity is a key target of ocean numerical climate models in general and, therefore, a key target for our ocean climate emulators.
The structure of the zonal mean of potential temperature (Figure~\ref{fig:fig1}b) is captured by the two emulators, demonstrating significant skill at reproducing the profile from OM4 (see S6 for salinity structure).
The average mean absolute error (MAE) is 5.7$\times {10^{-3}}~^\circ C$ for $\mathcal{F}_{\text{thermo+dynamic}}$ and 4.5$\times {10^{-3}}~^\circ C$ for $\mathcal{F}_{\text{thermo}}$, with a pattern correlation of roughly .99 for both emulators.
The outputs show a robust thermocline structure, subtropical gyres, and a region of North Atlantic deep water formation. 
However, both emulators in the northern hemisphere show too warm and too salty high latitudes (around 55N), too cold and too fresh mid-latitudes, and Arctic signals down to 750$\mathrm{m}$ depth (Figures S2 and S7).
The biases are consistent with underestimating the northward heat transport by the ocean. 
The potential temperature and salinity biases in the Southern Ocean for the $\mathcal{F}_{\text{thermo+dynamic}}$ emulator are reminiscent of residual transport changes, with opposite signed biases in the Southern Ocean and in the region north of it.
The $\mathcal{F}_{\text{thermo}}$ emulator is warmer than $\mathcal{F}_{\text{thermo+dynamic}}$, at most depths (Fig. S2). 

We performed several experiments to test the sensitivity of the emulators to different training choices.
The emulators' skill is unchanged when using different seeds and start dates, so the trained models are statistically reproducible.
We measure robustness by calculating the root mean square error (RMSE) of rollouts with 5 different seeds and rollouts initialized with ocean states taken 6 months apart. 
The RMSEs show little variance across the different trained models (Fig. \ref{fig:fig1}c). The standard deviation of the RMSEs across training seeds in the emulators $\mathcal{F}_{\text{thermo}}$ and $\mathcal{F}_{\text{thermo+dynamic}}$ are 0.0033 and 0.00225, respectively.

\begin{figure}
\centering
\includegraphics[width=\textwidth]{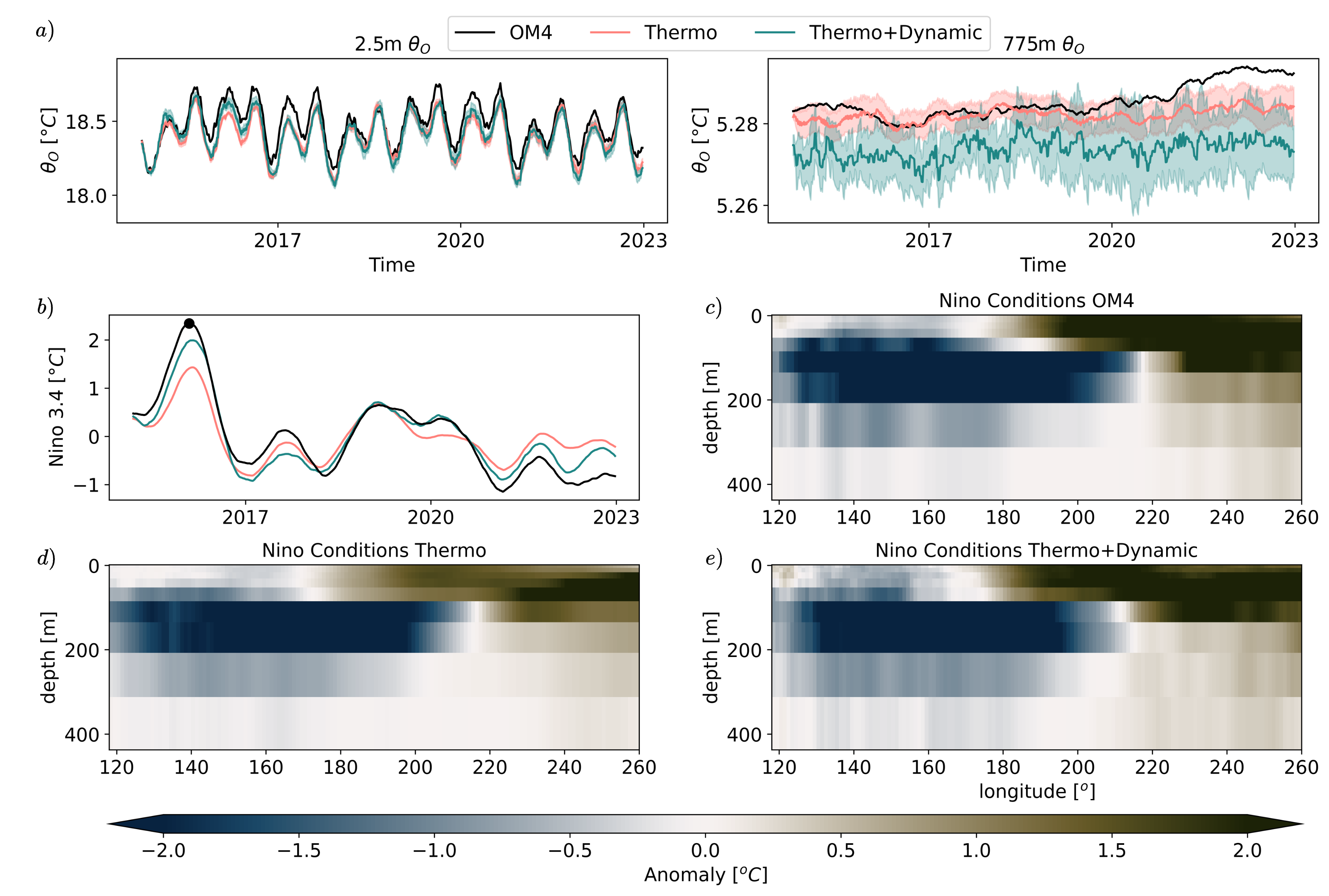}
\caption{a) Spatially averaged timeseries of potential temperature $\theta_O$ at depths 2.5$\mathrm{m}$ (left) and 775$\mathrm{m}$ (right) over the test set comparing the ground truth OM4 (black), and predictions from $\mathcal{F}_{\text{thermo}}$ (red) and $\mathcal{F}_{\text{thermo+dynamic}}$ (green).  The mean prediction and its variance (indicated by shading) are plotted over 5 initial seeds of training for each model. b) Nino 3.4 index timeseries over the test set for the ground truth (OM4, black) and predictions ($\mathcal{F}_{\text{thermo}}$, red;  $\mathcal{F}_{\text{thermo+dynamic}}$, green). Anomalies are averaged over rolling 150-day windows.
c-e) Meridionally averaged depth profile of potential temperature anomalies in the tropics during the peak Nino event (marked by a black dot in the timeseries) over the test set for OM4 (c), $\mathcal{F}_{\text{thermo}}$ (d) and $\mathcal{F}_{\text{thermo+dynamic}}$ (e). Anomalies in (c)-(e) are averaged over a 15-day window. } 
\label{fig:fig2}
\end{figure}

The potential-temperature  timeseries at 2.5$\mathrm{m}$ and 775$\mathrm{m}$ (Figure~\ref{fig:fig2}a) are further indicators that both emulators capture the climatological means and the upper ocean response to variable atmospheric forcing. 
The standard deviation of the 2.5$\mathrm{m}$ potential temperature for OM4, and the emulators $\mathcal{F}_{\text{thermo}}$ and $\mathcal{F}_{\text{thermo+dynamic}}$ are 6.8$\times {10^{-2}}~ ^{\circ} C$, 4.35$\times {10^{-2}}~^{\circ} C$ and 5.26$\times {10^{-2}}~^{\circ} C$ respectively, while the standard deviations of the 775$\mathrm{m}$ potential temperature are 2.3 $\times {10^{-3}}~^{\circ} C$, 1.0$ \times {10^{-3}}~^{\circ} C$ and 2.1$\times {10^{-3}}~^{\circ} C$, respectively. The emulators capture a large portion of the variability, but with some biases (Fig.~\ref{fig:fig2}b). The standard deviations are calculated after removing both the trend and the climatology from the timeseries (See Figure S8 for additional timeseries of potential temperature, along with salinity, zonal velocity, and meridional velocity, and Figure S10 for bias maps).

The emulator can skillfully emulate El Niño-Southern Oscillation (ENSO) response in both warm and cold phases (Figure~\ref{fig:fig2}b) and S11).
The smallest fluctuations in the Nino 3.4 timeseries are the hardest for the emulators to capture.
The emulator responses are in phase with OM4 for all years shown, but the amplitude is altered. 
$\mathcal{F}_{\text{thermo+dynamic}}$ exhibits higher skill than $\mathcal{F}_{\text{thermo}}$ in capturing the magnitude of ENSO events. 
We hypothesized that providing the velocities, whose data contain shorter time-scales and larger variability, helps the emulator produce larger ENSO events.
$\mathcal{F}_{\text{thermo}}$ still manages to detect the correct phase and structure (Figure ~\ref{fig:fig2} b,d) despite producing events with smaller magnitudes, both at the surface and in the upper ocean. 
The emulators capture the deepening and shoaling of the equatorial thermocline from equatorial Kelvin waves for the strongest events (Figure ~\ref{fig:fig2} d, e). 
The magnitude of subsurface anomalies for the emulators is weaker than for OM4. For the Nino 3.4 timeseries (Figure ~\ref{fig:fig2} b), the MAE is 0.0077$~^\circ C$ for $\mathcal{F}_{\text{thermo+dynamic}}$ and 0.0124$~^\circ C$ for $\mathcal{F}_{\text{thermo}}$, with correlations of 0.905 and 0.7017, respectively. For the ENSO profiles (Figure ~\ref{fig:fig2} (c)-(e)), the MAE is 0.01$~^\circ C$ and 0.07$~^\circ C$ for the emulators $\mathcal{F}_{\text{thermo+dynamic}}$ and $\mathcal{F}_{\text{thermo}}$ respectively, and their pattern correlations are 0.976 and 0.973, respectively.

For the ocean emulator $\mathcal{F}_{\text{thermo+dynamic}}$ that uses all variables, we noticed that the potential temperature and salinity fields exhibit atypically high spatial variability, with scales more characteristic of velocity so we posit that this results from using velocity inputs. See Figures S16-S17 for maps of variability for our emulators. 
This result is consistent with \citeA{subel2024building}.
We hypothesize that this may arise from the large separation in timescales and variability between velocity and potential temperature in the ocean. 

Finally, despite capturing the mean and climatology of ocean variables, the emulators struggle to capture the magnitude of the small, but systematic potential temperature trends (Figure S1 global mean $10^{-3} ~^{\circ} C/yr$) over the same 8-year period (Figure~\ref{fig:fig2}a and S1, S3); for most depths the trained models underestimate trends by 20\% to 50\% relative to OM4. 
Of the two emulators, $\mathcal{F}_{\text{thermo}}$ has higher skill in capturing the global heat changes (Figure S9). 
The salinity trends in OM4 are weak, due to the small forcing, and to the use of salinity restoring boundary conditions. 
For both emulators, the trends are 7-8 orders of magnitude less than the mean value, consistent with the numerical representation of variables within the learned models, suggesting that the models conserve properties of the OM4 data although strict conservation is not imposed (Figures S4-S5). 



\subsection{Long-term stability}
\label{sec:Long-term}

We also evaluated, the ability of the emulators to produce long control experiments, without retraining.
For these experiments, we use repeat boundary conditions over 10 years (described in Section \ref{sec:Evaluation}) chosen to contribute a near-zero net heat flux, allowing the emulators to run for arbitrarily long periods of time while minimizing potential temperature drift. 

\begin{figure}
\noindent\includegraphics[width=\textwidth]{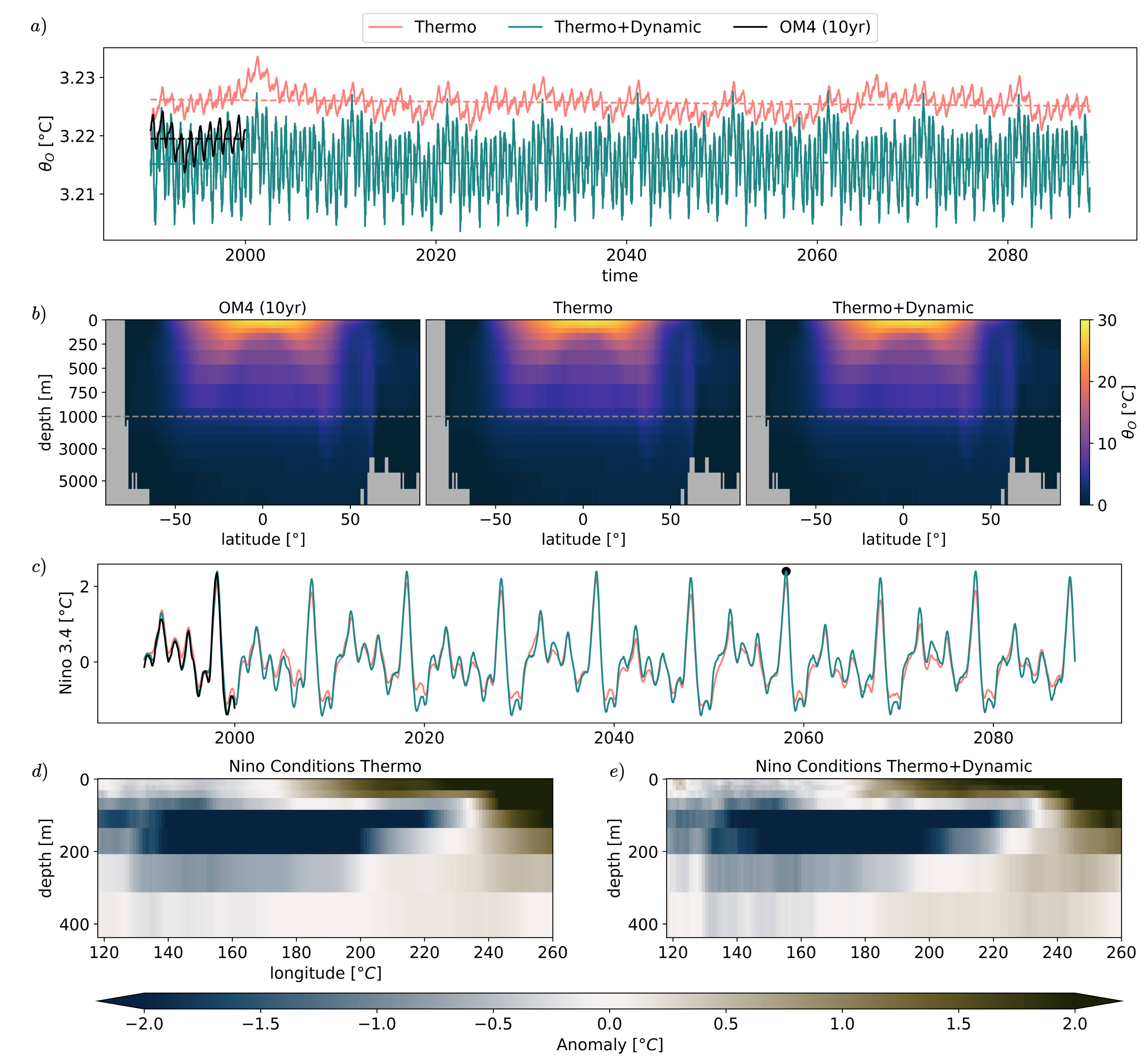}
\caption{a) Globally averaged potential temperature ($\theta_O$) timeseries over a 100-year control run, comparing the 10-year ground truth OM4 (black) and predictions from $\mathcal{F}_{\text{thermo}}$ (red) and  $\mathcal{F}_{\text{thermo+dynamic}}$ (green). b) Time-averaged potential temperature ($\theta_O$) depth profile over a 100-year control run, comparing the 10-year ground truth OM4 (left) and predictions from $\mathcal{F}_{\text{thermo}}$ (middle) and $\mathcal{F}_{\text{thermo+dynamic}}$ (right). c) Nino 3.4 index timeseries over a 100-year control run, comparing the 10-year repeat for the ground truth (OM4, black) and predictions ($\mathcal{F}_{\text{thermo}}$, red;  $\mathcal{F}_{\text{thermo+dynamic}}$, green).
d-e) Meridionally averaged depth profile of potential temperature anomalies in the tropics during the peak Nino event (marked by a black dot in the timeseries) over the test set for $\mathcal{F}_{\text{thermo}}$ (d) and $\mathcal{F}_{\text{thermo+dynamic}}$ (e). Anomalies are as in Fig.~\ref{fig:fig2}.}
\label{fig:fig3}
\end{figure}

Both emulators converge to an equilibrium, maintaining a global mean potential temperature close to OM4 throughout a century of integration (Figure \ref{fig:fig3}a).
The global mean temperatures are 3.225 $~^{\circ} C/yr$ for $\mathcal{F}_{\text{thermo}}$ and 3.215 $~^{\circ} C/yr$ for $\mathcal{F}_{\text{thermo+dynamic}}$, compared to 3.219 $~^{\circ} C/yr$ for OM4.
In addition, $\mathcal{F}_{\text{thermo+dynamic}}$ over-predicts the variability in potential temperature, likely extrapolating some fast dynamics via the velocities variables.
This issue is exacerbated in the deeper layers of the ocean, which have little variability in the original dataset. 
The temperature structure is again well preserved for the long rollouts (Figure \ref{fig:fig3}b), with different structures in potential temperature biases (S12) than for the 8-year test data (S2). 

We examine the emulators'  skill in reproducing variability over these long timescales.
Since we are reusing the same 10-year cycle to drive the emulator, we expected some persistent features to appear when looking at a phenomenon such as the response to ENSO.
Although both emulators can produce appropriate Nino 3.4 anomalies for the entire century rollout (Figure \ref{fig:fig3}c) and S13), $\mathcal{F}_{\text{thermo+dynamic}}$ shows stronger peak-to-peak amplitude, but little cycle-to-cycle variability - perhaps due to the strong coupling of velocity with the wind stress forcing, whereas $\mathcal{F}_{\text{thermo}}$ shows more aperiodic variability across years. 

To further test stability, we generate a 400-year rollout, with an identical forcing setup as for the century-long run. 
Both emulators remain stable (Figure S15). $\mathcal{F}_{\text{thermo}}$ has the added benefit of exhibiting long-term aperiodic variability in potential temperature and salinity, despite the repeat forcing, across the centuries. The long experiments were reproduced using a repeat forcing period from the test set i.e. 2014-2022, producing similar results (Figure S19).

\section{Discussion}
\label{sec:discussion}

We produce a computationally cheap machine-learning (ML) emulator of a state-of-the-art ocean model, namely OM4 \cite{adcroft_2019}.
The ML architecture consists of a modified ConvNeXt UNet  \cite{dheeshjith2024transfer}. 
The reduced order model -- \textit{Samudra} -- predicts key ocean variables, sea surface height, temperature, and salinity, across the full depth of the world oceans while remaining stable for centuries. 
Integrating OM4 for 100 years takes approximately 8 days using 4,671 CPU cores, whereas our fastest (thermo) emulator completes the same task in about 1.3 hours on a single 40GB A100 GPU. 
This represents approximately a 150x increase in SYPD (simulated years per day) for Samudra compared to OM4. Some of this speed up can be attributed to Samudra: i) using a 5 day time step (vs. 15 minutes in OM4); ii) operating on a spatially coarser grid. However, we note that Samudra makes 
predictions with the implicit spatial skill of the finer resolution OM4, whereas existing coarser GCMs with eddy parameterization tend to show worse biases (e.g. see fig. 9 of \citeA{adcroft_2019} for a 1/2-degree GCM).

The emulator performs well on a range of metrics related to the model climatology and its variability on the test set and long control simulations. 
The emulator produces accurate climatologies over the last 8 years of the OM4 simulations and is robust to changes in seeds and initial conditions. 
Furthermore, it can capture variability (e.g., ENSO response to forcing). 
Therefore, these emulators could be used to study the contemporary ocean and climate at a significant reduction in cost compared to OM4.

The emulator, however, struggles to capture trends under a range of surface heat flux forcings (see Supporting Information), similarly to the surface emulators in \citeA{dheeshjith2024transfer}.
We performed idealized forced experiments using the same repeated atmospheric forcing generated for the control experiment and a spatially uniform linear forcing of varying magnitudes for the surface heat flux. 
Figure S16 showcases the ocean heat content trends predicted by $\mathcal{F}_{\text{thermo}}$ under linear surface heat flux increases of 1, 0.5, 0.25, and 0 $W/m^2$. 
The patterns of ocean heat uptake are reminiscent of ocean-only and coupled forced numerical experiments \cite{todd2020ocean,couldrey2020causes}, with dipole patterns in the Southern Ocean and North Atlantic sinking region (Figure S14). However, the magnitude of change is too weak compared to the forcing (Figure S16). 
Similar behavior of weak generalization under climate change is also observed in the atmosphere climate emulator, ACE \cite{watt2023ace}, but improved when a slab ocean model is added \cite{clark2024ace2}.  

Here, we could not produce an emulator that simultaneously captures the trends in the test data and remain stable for centuries. Further work is needed to explore the reasons for the issues and would require new numerical simulations. 

The lack of generalization reflected in the weak warming trends could be due to the training data. 
The effects of an initial drift can be alleviated by pruning years 1958 to 1975 from the training data, which removes the bulk of this adjustment period. 
Yet, different depths and regions adjust more slowly, and some of this continued adjustment may remain in the data since the time scale of equilibration of the model is hundreds of years.  
Another reason for the trend bias could be the forcing datasets.
The atmospheric forcing imposed on the ocean implicitly results from the real ocean-atmosphere coupling. 
Therefore, the atmospheric forcing has felt a changing ocean circulation, particularly in the North Atlantic \cite{chemke2020identifying}. 
The resulting effect is that the ``forcing" applied to the ocean emulator is not entirely decoupled from the ocean response, potentially leading to some biases in the response, as in \citeA{todd2020ocean,couldrey2020causes,zanna2019global}. 
We alleviated these issues by adding an extra forcing input, namely the cumulative heat forcing, which led to a more skillful model capable of capturing the global warming trend. However, this model was unstable under climate-change forcing past 50 years. 
Alternatively, it is possible that learning to predict the model state directly may not be optimal. We explored learning tendencies,  
which improved performance for the warming trends but, again, was unstable over long timescales.
A challenge going forward is designing faithful emulators capable of capturing trends while remaining stable in long rollouts. 

Despite the limited response to future climate forcing, Samudra is skillful at emulating the contemporary ocean and is therefore an affordable emulation of expensive ocean circulation models. Without further modification, Samudra could be used in studies requiring large ensembles (e.g., uncertainty quantification, extreme events) or to enhance and accelerate operational applications (e.g., data assimilation). More opportunities emerge if we consider refining training for Samudra, e.g., to revised versions of OM4 or to other models, which could greatly accelerate climate model development by allowing evaluations of long, yet affordable, rollouts. This includes coupling Samudra with ACE \cite{watt2023ace} to emulate CM4. 

\section*{Open Research Section}

The code for training the models along with generating rollouts and plots is available on GitHub at \url{https://github.com/m2lines/Samudra}, while the model weights and data are hosted on Hugging Face at  \url{https://huggingface.co/M2LInES/Samudra} and \url{https://huggingface.co/datasets/M2LInES/Samudra-OM4}, respectively. Additionally, data from \citeA{cisl_rda_dsd277006} was also used in the Supporting Information. The code is also version tagged and archived at \citeA{dheeshjith-doi-software} via zenodo.

\acknowledgments
This research received support through Schmidt Sciences, LLC, under the M$^2$LInES project. 
We thank all members of the M$^2$LInES team for helpful discussions and their support throughout this project. We gratefully acknowledge Karthik Kashinath and the NVIDIA team for providing us access to NERSC resources, which were instrumental in supporting this work.
This research was also supported in part through the NYU IT High Performance Computing resources, services, and staff expertise. We also thank the reviewers for their useful comments.

\nocite{cisl_rda_dsd277006}

\bibliography{references}
\newpage
\section*{Supporting Information}




\noindent\textbf{Text S1.}
Here we describe how we calculated $\operatorname{Q}_{anom}$. 


\begin{equation}
    \operatorname{Q}_{anom}(t,y,x) = \operatorname{Q}(t,y,x) - Clim(\operatorname{Q})(t,y,x)
\end{equation}
where Clim is the climatology of $ \operatorname{Q}$ over the entire data.

\noindent\textbf{Text S2.} Calculation of Metrics

Consider a predicted ocean state $\boldsymbol{\tilde \Phi}_{t}^{[j,k,l]}$, its corresponding ground truth state $\boldsymbol{\Phi}_{t}^{[j,k,l]}$ at time $t$, channel $j$, latitude $k$ and longitude $l$, and the normalized volume $V(j,k,l)$ at channel $j$, latitude $k$ and longitude $l$. 

\begin{equation}
RMSE(\boldsymbol{\tilde \Phi}, \boldsymbol{\Phi}) = \frac{1}{T}\sum_{t}\sqrt{\sum_{j,k,l}V(j,k,l)\bigg(\boldsymbol{\tilde \Phi}_{t}^{[j,k,l]} - \boldsymbol{\Phi}_{t}^{[j,k,l]}\bigg)^2}
\end{equation}

\begin{equation}
MAE(\boldsymbol{\tilde \Phi}, \boldsymbol{\Phi}) = \frac{1}{T}\sum_{t} \left|\sum_{j,k,l}V(j,k,l)\bigg(\boldsymbol{\tilde \Phi}_{t}^{[j,k,l]} - \boldsymbol{\Phi}_{t}^{[j,k,l]}\bigg)\right|
\end{equation}

\begin{equation}
Corr(\boldsymbol{\tilde \Phi}, \boldsymbol{\Phi}) = \frac{1}{T}\sum_{t} \frac{\sum_{j,k,l}V(j,k,l) \boldsymbol{\tilde \Phi}_{t}^{[j,k,l]}\boldsymbol{\Phi}_{t}^{[j,k,l]}}{\sqrt{\sum_{j,k,l}V(j,k,l) \big(\boldsymbol{\tilde \Phi}_{t}^{[j,k,l]}\big)^2 \sum_{j,k,l}V(j,k,l) \big(\boldsymbol{\Phi}_{t}^{[j,k,l]}\big)^2}}
\end{equation}

where $T$ is the time period over which we calculate the metrics.

\newpage
\renewcommand{\thefigure}{S\arabic{figure}}
\setcounter{figure}{0} 
\begin{figure}
    \centering
    \includegraphics[width=\linewidth]{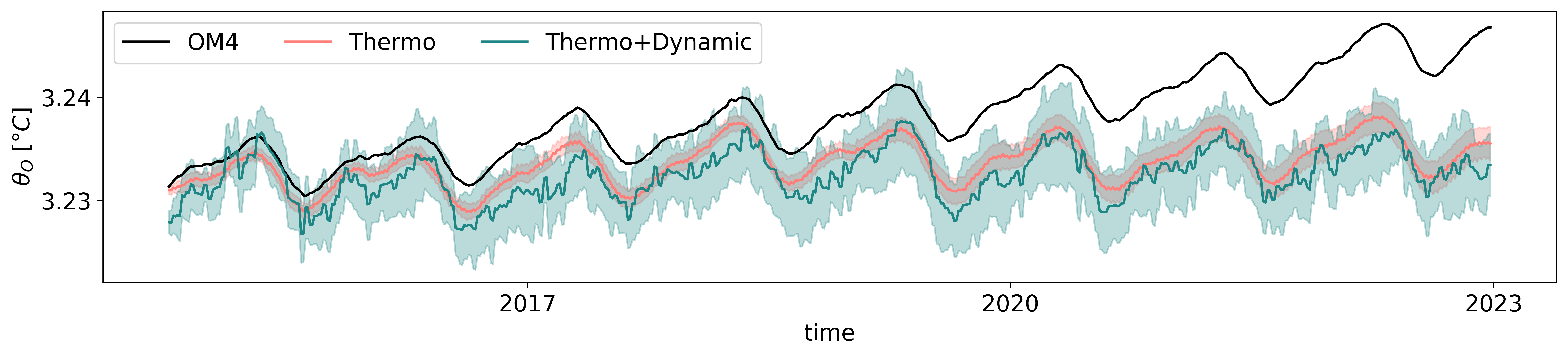}
    \caption{Spatially averaged potential temperature ($\theta_O$) time series over an 8-year test set comparing the ground truth OM4 (black), and predictions from $\mathcal{F}_{\text{thermo}}$ (red), and $\mathcal{F}_{\text{thermo+dynamic}}$ (green). The mean prediction and its variance (indicated by shading) are plotted over 5 initial seeds of training for each model.}
    \label{fig:OM4HC}
\end{figure}

\begin{figure}
    \centering
    \includegraphics[width=\linewidth]{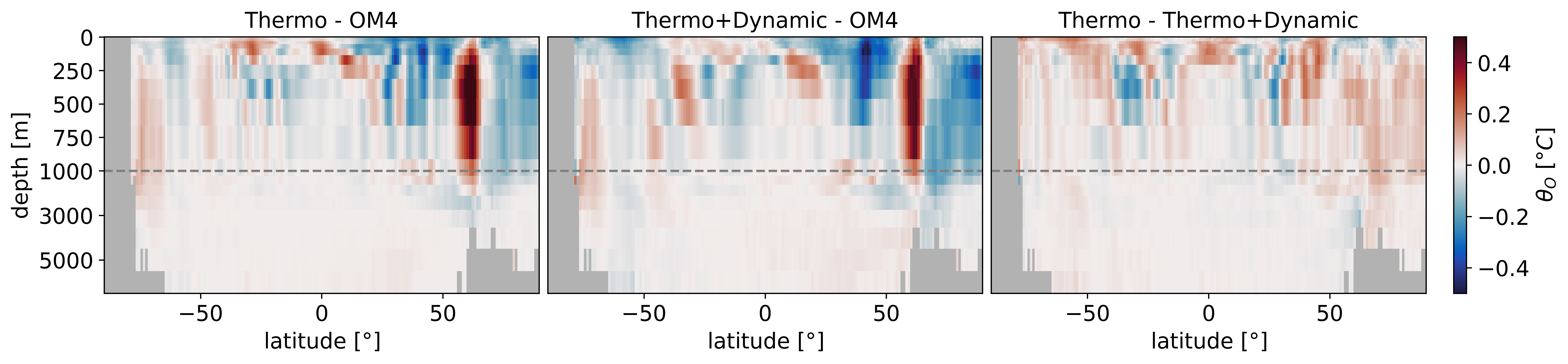}
    \caption{Time- and zonally-averaged potential temperature ($\theta_O$) biases (relative to OM4) for an 8-year test set: $\mathcal{F}_{\text{thermo}}$ (left), $\mathcal{F}_{\text{thermo+dynamic}}$ (center), and the difference between $\mathcal{F}_{\text{thermo}}$ and $\mathcal{F}_{\text{thermo+dynamic}}$ (right).}
    \label{fig:OM4HC}
\end{figure}

\begin{figure}
    \centering
    \includegraphics[width=\linewidth]{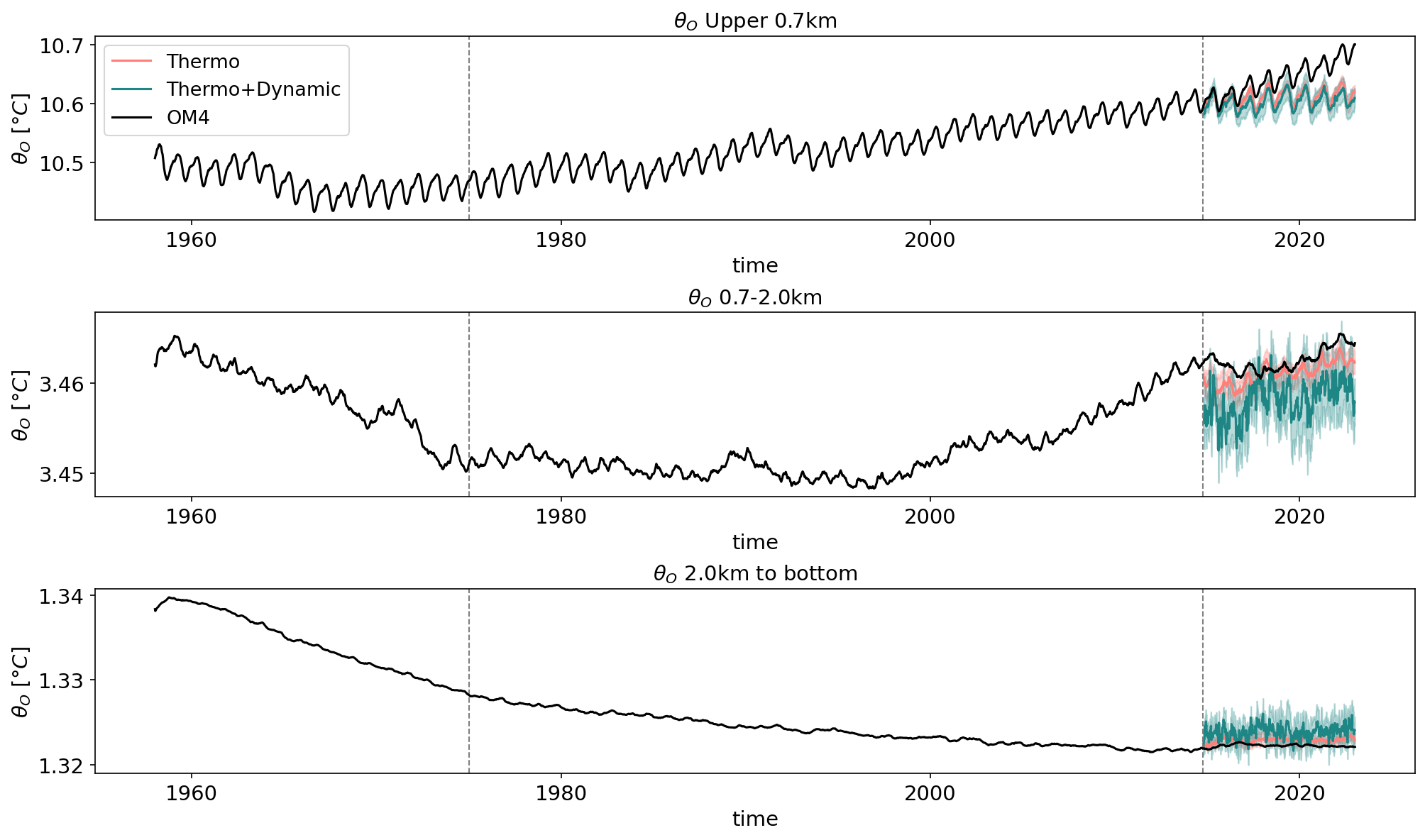}
    \caption{Spatially averaged potential temperature ($\theta_O$) trends of the entire ground truth data OM4 (black), and 8-year test set predictions from $\mathcal{F}_{\text{thermo}}$ (red) and $\mathcal{F}_{\text{thermo+dynamic}}$ (green) at depth levels 0–700 m, 700–2000 m, and 2000–6000 m. Vertical lines indicate the section of training data considered. The mean prediction and its variance (indicated by shading) are plotted over 5 initial seeds of training for each model.}
    \label{fig:OM4HC}
\end{figure}

\begin{figure}
    \centering
    \includegraphics[width=\linewidth]{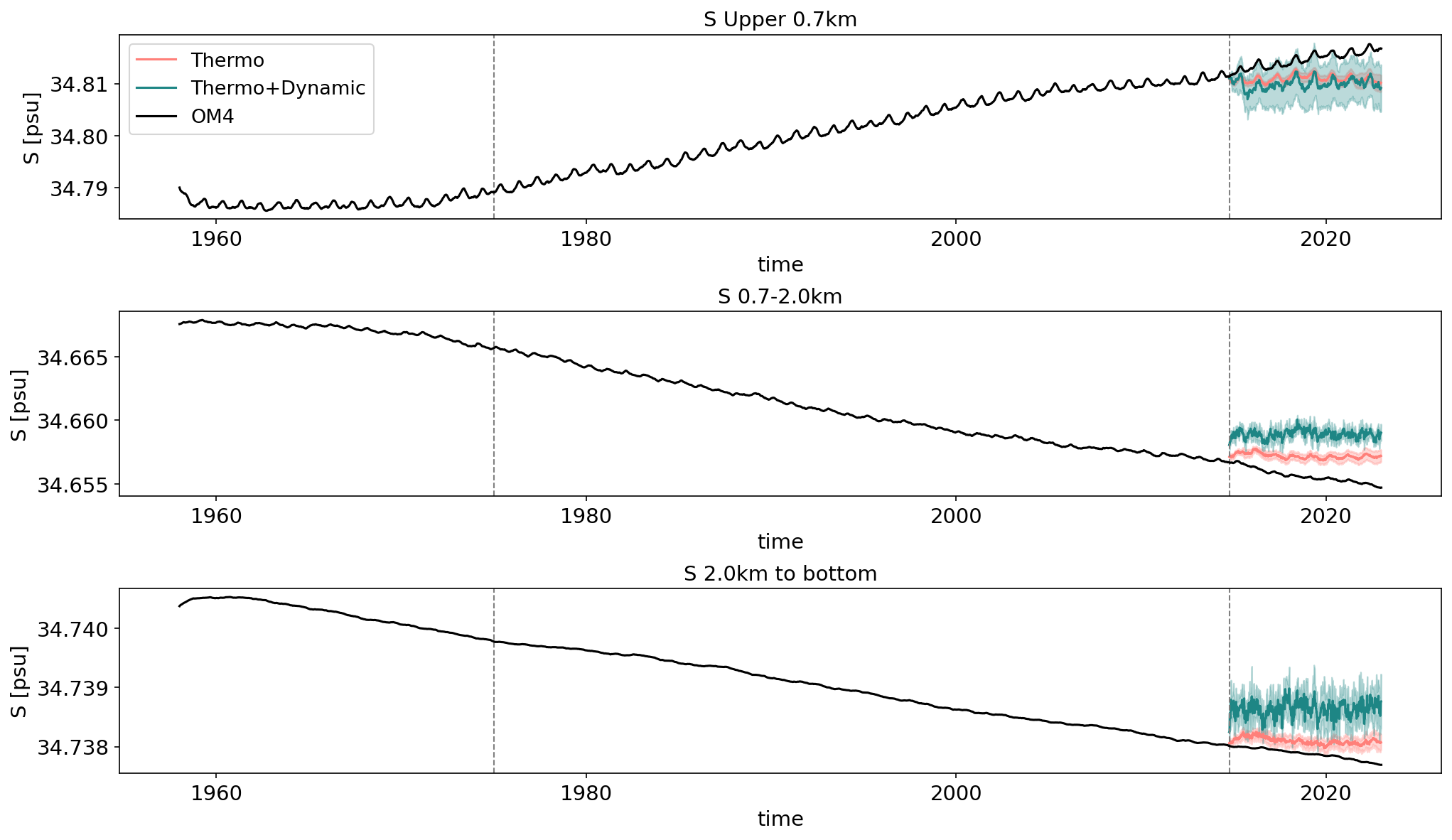}
    \caption{Spatially averaged salinity ($S$) trends of the entire ground data truth OM4 (black), and 8-year test set predictions from $\mathcal{F}_{\text{thermo}}$ (red) and $\mathcal{F}_{\text{thermo+dynamic}}$ (green) at depth levels 0–700 m, 700–2000 m, and 2000–6000 m. Vertical lines indicate the section of training data considered. The mean prediction and its variance (indicated by shading) are plotted over 5 initial seeds of training for each model.}
    \label{fig:OM4HC}
\end{figure}

\begin{figure}
    \centering
    \includegraphics[width=\linewidth]{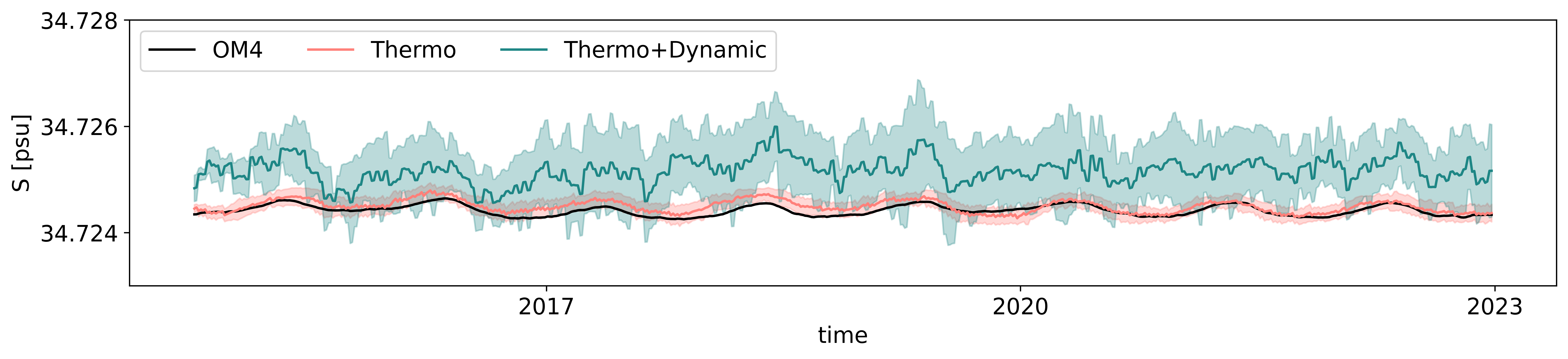}
    \caption{Spatially averaged salinity ($S$) time series over an 8-year test set comparing the ground truth OM4 (black), and predictions from $\mathcal{F}_{\text{thermo}}$ (red), and $\mathcal{F}_{\text{thermo+dynamic}}$ (green). The mean prediction and its variance (indicated by shading) are plotted over 5 initial seeds of training for each model.}
    \label{fig:OM4HC}
\end{figure}

\begin{figure}
    \centering
    \includegraphics[width=\linewidth]{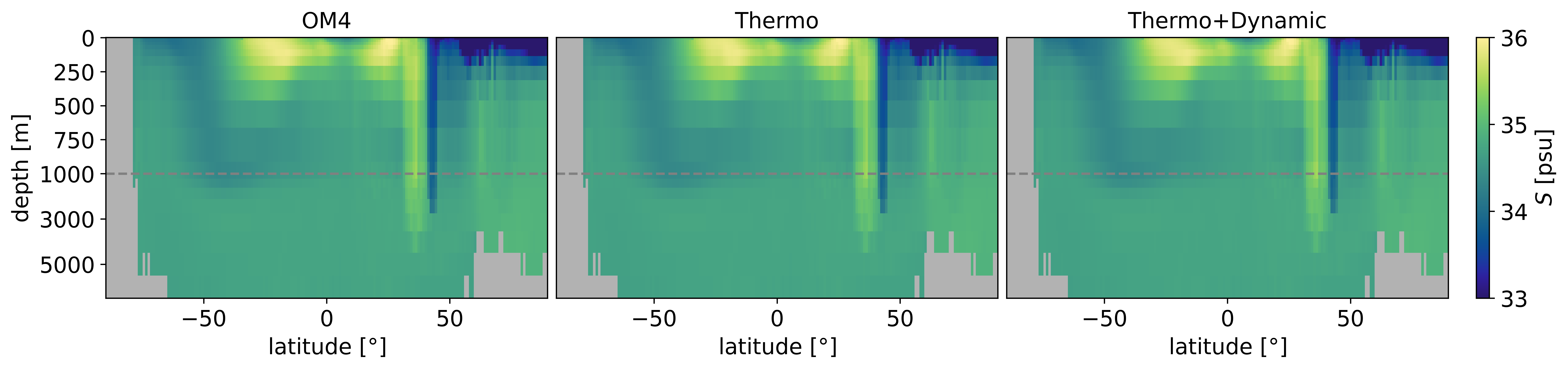}
    \caption{Time- and zonally-averaged salinity ($S$) for an 8-year test set:  ground truth OM4 (left), $\mathcal{F}_{\text{thermo}}$ (center), and $\mathcal{F}_{\text{thermo+dynamic}}$ (right).}
    \label{fig:OM4HC}
\end{figure}

\begin{figure}
    \centering
    \includegraphics[width=\linewidth]{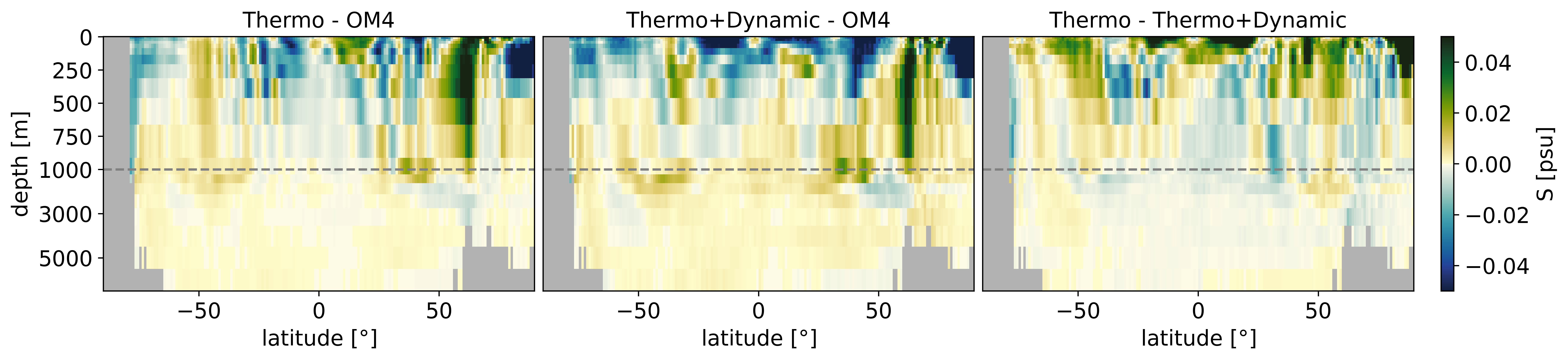}
    \caption{Time- and zonally-averaged salinity ($S$) biases (relative to OM4) for $\mathcal{F}_{\text{thermo}}$ (left), $\mathcal{F}_{\text{thermo+dynamic}}$ (center), and the difference between $\mathcal{F}_{\text{thermo}}$ and $\mathcal{F}_{\text{thermo+dynamic}}$ (right) for an 8-year test set.}
    \label{fig:OM4HC}
\end{figure}

\begin{figure}
    \centering
    \includegraphics[width=\linewidth]{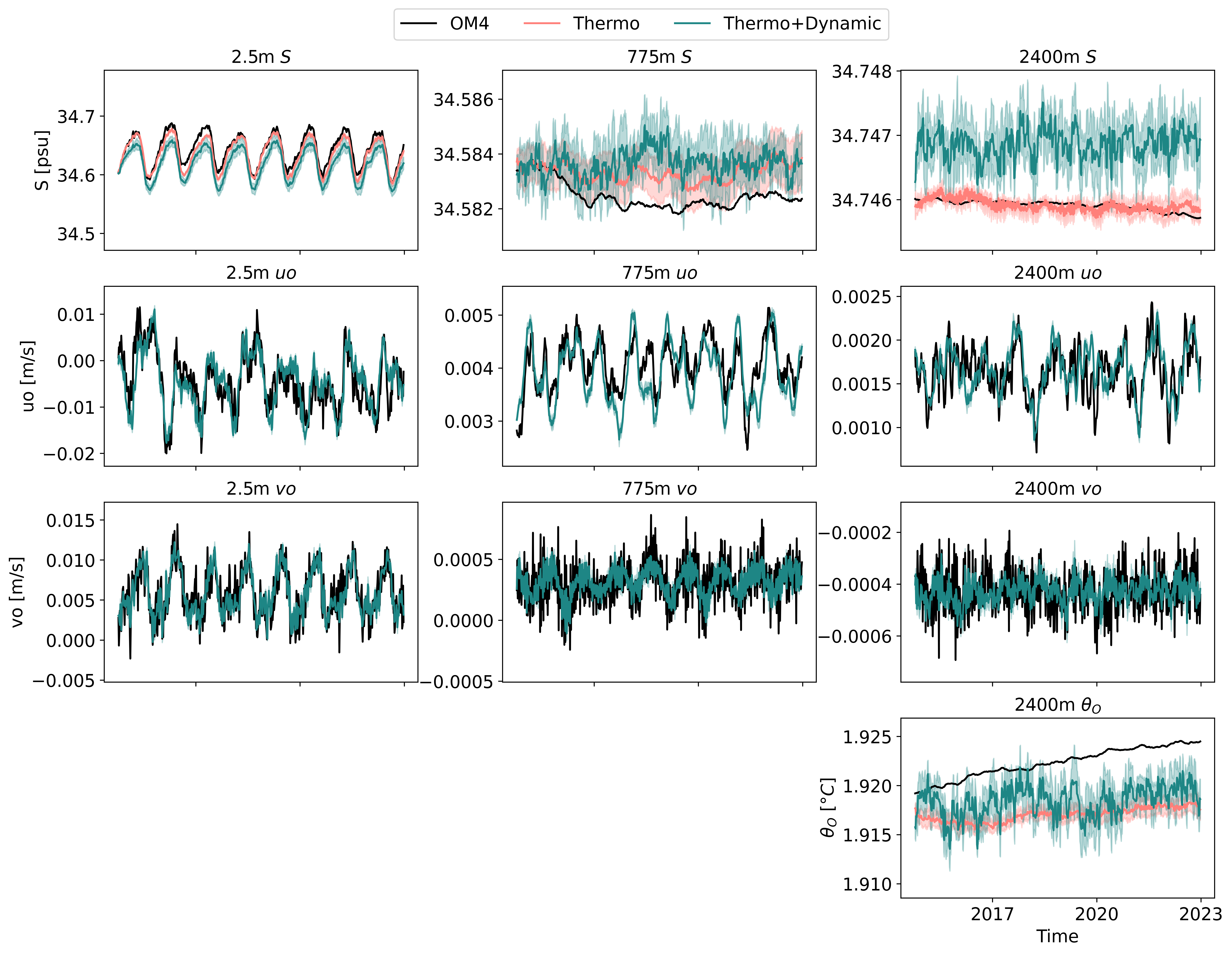}
    \caption{Spatially averaged time series over an 8-year test set for the ground truth OM4 (black), $\mathcal{F}_{\text{thermo}}$ (red), and $\mathcal{F}_{\text{thermo+dynamic}}$ (green). The first, second, and third rows correspond to salinity ($S$), zonal velocity ($uo$), and meridional velocity ($vo$) at depths of 2.5$\mathrm{m}$, 775$\mathrm{m}$, and 2400$\mathrm{m}$, respectively. The final plot in the bottom row represents potential temperature ($\theta_O$) at 2400$\mathrm{m}$. The mean prediction and its variance (indicated by shading) are plotted over 5 initial seeds of training for each model.}
    \label{fig:OM4HC}
\end{figure}

\begin{figure}
    \centering
    \includegraphics[width=\linewidth]{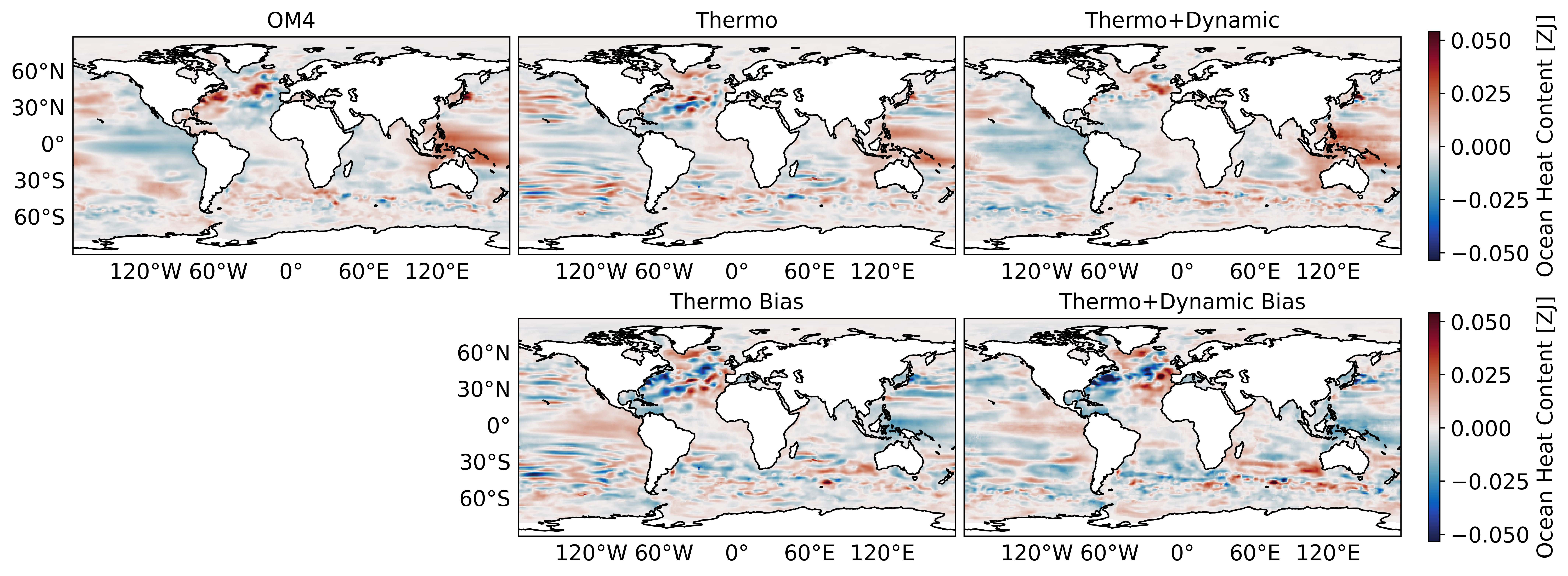}
    \caption{Global maps of Ocean Heat Content (OHC) evaluated over an 8-year test set, displaying the difference between the last and first year for the ground truth OM4 (top left), $\mathcal{F}_{\text{thermo}}$ (top center), and $\mathcal{F}_{\text{thermo+dynamic}}$ (top right). The corresponding bias maps are shown in the bottom row. }
    \label{fig:OM4HC}
\end{figure}

\begin{figure}
    \centering
    \includegraphics[width=\linewidth]{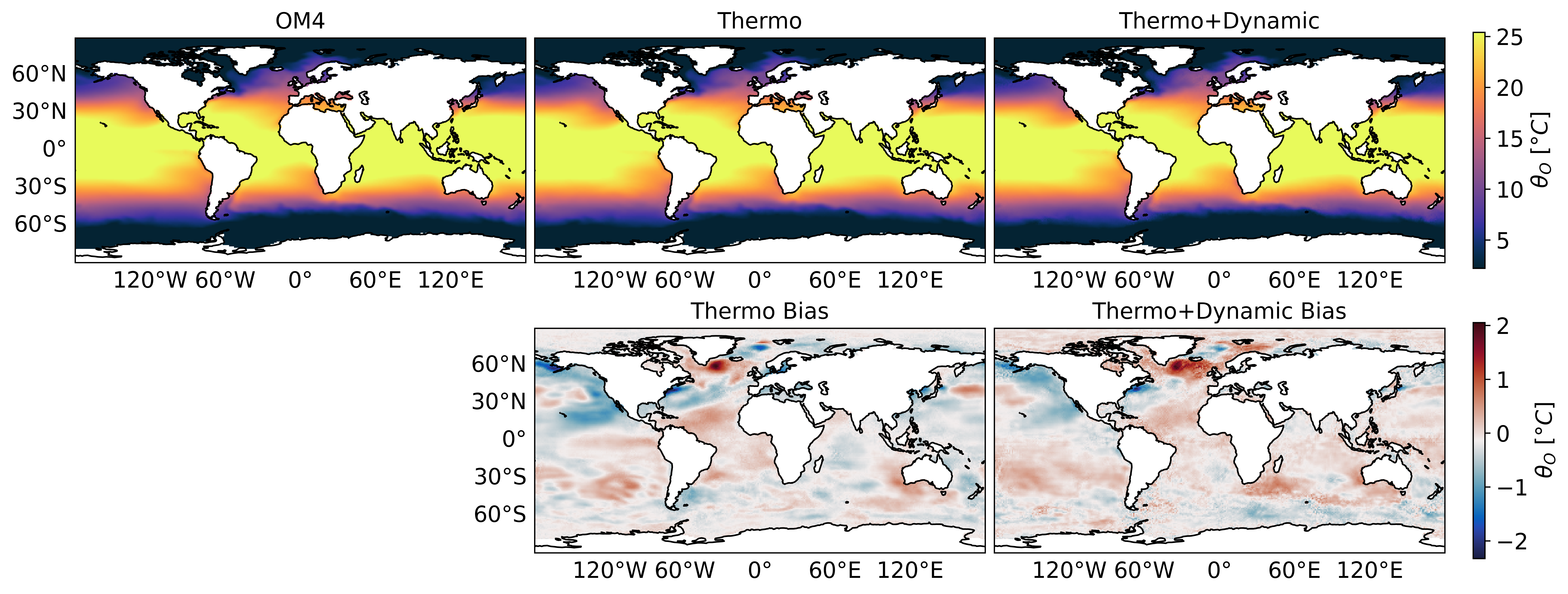}
    \caption{Time-averaged global maps of 2.5$\mathrm{m}$ potential temperature ($\theta_O$) evaluated over an 8-year test set for the ground truth OM4 (top left), $\mathcal{F}_{\text{thermo}}$ (top center), and $\mathcal{F}_{\text{thermo+dynamic}}$ (top right), with corresponding bias maps displayed in the bottom row.}
    \label{fig:OM4HC}
\end{figure}

\begin{figure}
    \centering
    \includegraphics[width=\linewidth]{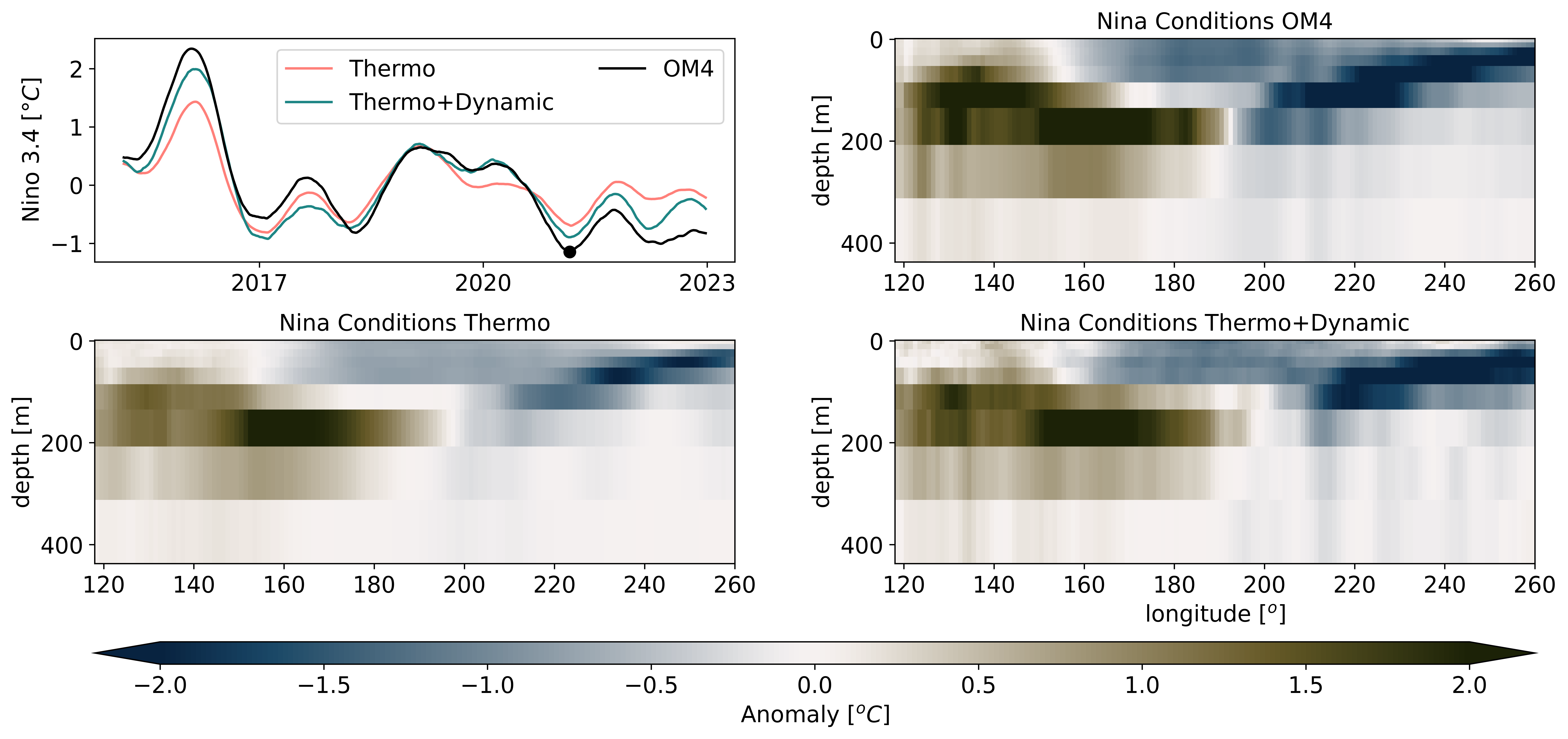}
    \caption{Timeseries of Nino 3.4 index over an 8-year test set, comparing the ground truth OM4 (black) with predictions from $\mathcal{F}_{\text{thermo}}$ (red) and $\mathcal{F}_{\text{thermo+dynamic}}$ (green). Here, we consider the 2.5$\mathrm{m}$ temperature anomalies. Anomalies are calculated relative to the 8-year climatology of OM4 and each emulator. Additionally, the depth structure of anomalies is shown for the peak Nina event (marked by a black dot in the timeseries). Anomalies are averaged over rolling 150-day windows in the timeseries while the anomalies in the depth structures are averaged over a 15-day (3-snapshot) window to reduce mesoscale variability.}
    \label{fig:OM4HC}
\end{figure}

\begin{figure}
    \centering
    \includegraphics[width=\linewidth]{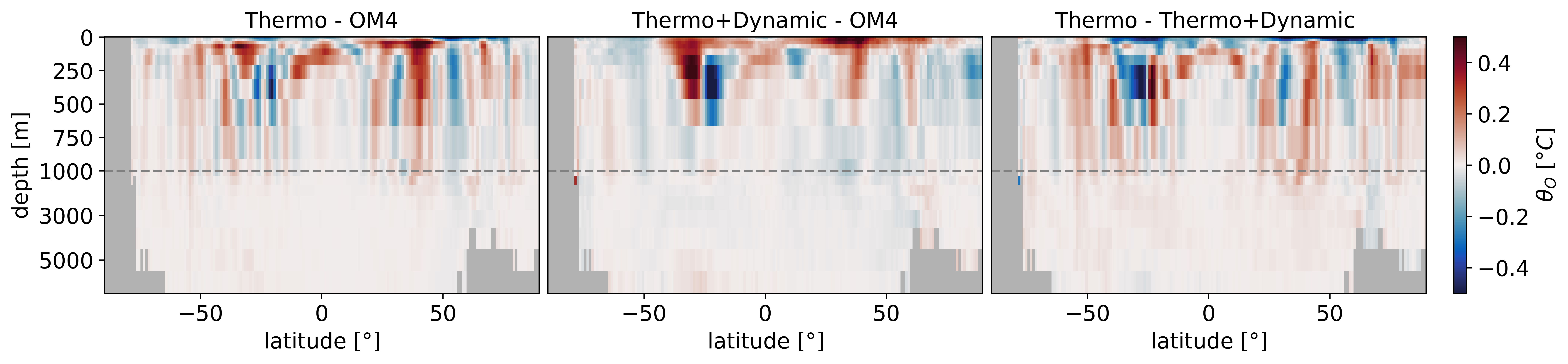}
    \caption{Time- and zonally-averaged potential temperature ($\theta_O$) biases (relative to OM4) for a 100-year control run forced with repeated atmospheric conditions taken from 1990-2000: $\mathcal{F}_{\text{thermo}}$ (left), $\mathcal{F}_{\text{thermo+dynamic}}$ (center), and the difference between $\mathcal{F}_{\text{thermo}}$ and $\mathcal{F}_{\text{thermo+dynamic}}$ (right). We compare the average for the 10-year period (1990–2000) of OM4 with the average of the 100-year emulator run.}
    \label{fig:OM4HC}
\end{figure}

\begin{figure}
    \centering
    \includegraphics[width=\linewidth]{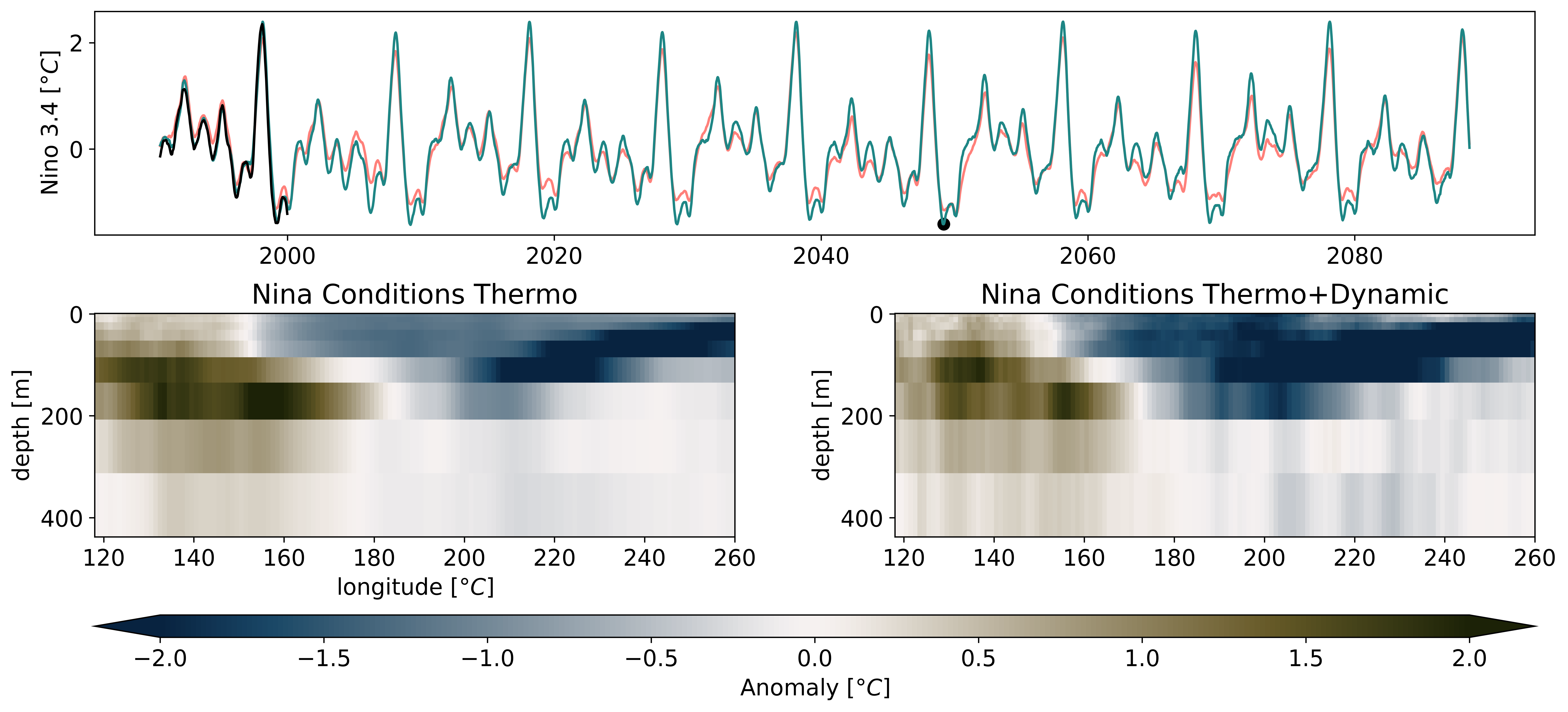}
    \caption{Timeseries of Nino 3.4 index over a 100-year control run, comparing the 10-year repeat ground truth OM4 (black) with predictions from $\mathcal{F}_{\text{thermo}}$ (red) and $\mathcal{F}_{\text{thermo+dynamic}}$ (green). Here, we consider the 2.5$\mathrm{m}$ temperature anomalies. Anomalies are calculated relative to the 10-year climatology of OM4 and 100-year climatology of each emulator. Additionally, the depth structure of anomalies is shown for the peak Nina event (marked by a black dot in the timeseries).  Anomalies are averaged over rolling 150-day windows in the timeseries while the anomalies in the depth structures are averaged over a 15-day (3-snapshot) window to reduce mesoscale variability.}
    \label{fig:OM4HC}
\end{figure}

\begin{figure}
    \centering
    \includegraphics[width=\linewidth]{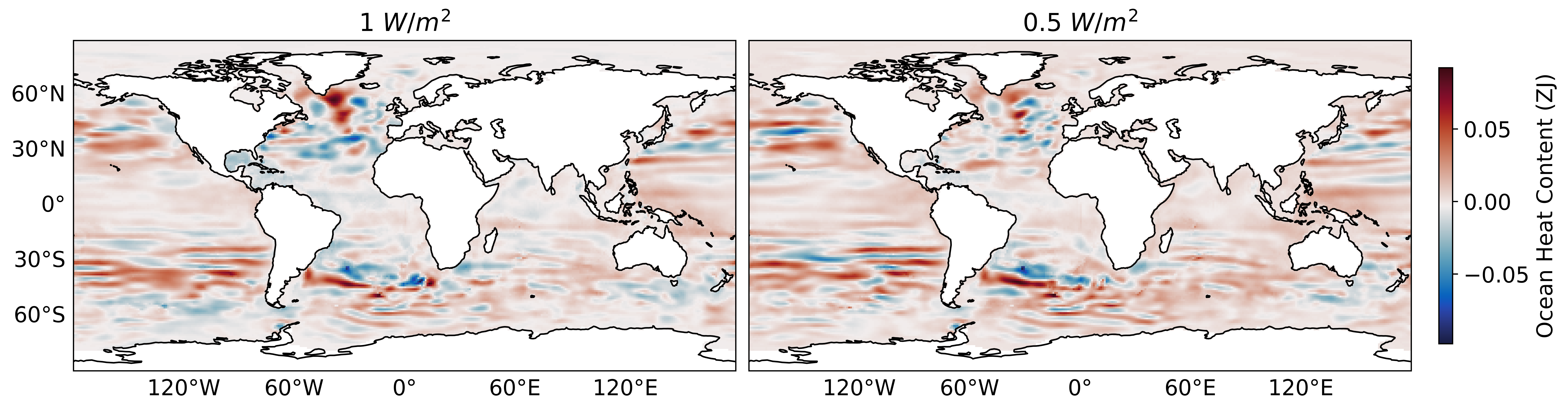}
    \caption{OHC Global Maps for the $\mathcal{F}_{\text{thermo}}$ emulator, evaluated over a 100-year climate run forced with $1W/m^2$ (left) and $0.5W/m^2$ (right) yearly increase in global heat flux forcing, showing the difference between the time-averaged last 5 years and first 5 years.}
    \label{fig:OM4HC}
\end{figure}

\begin{figure}
    \centering
    \includegraphics[width=\linewidth]{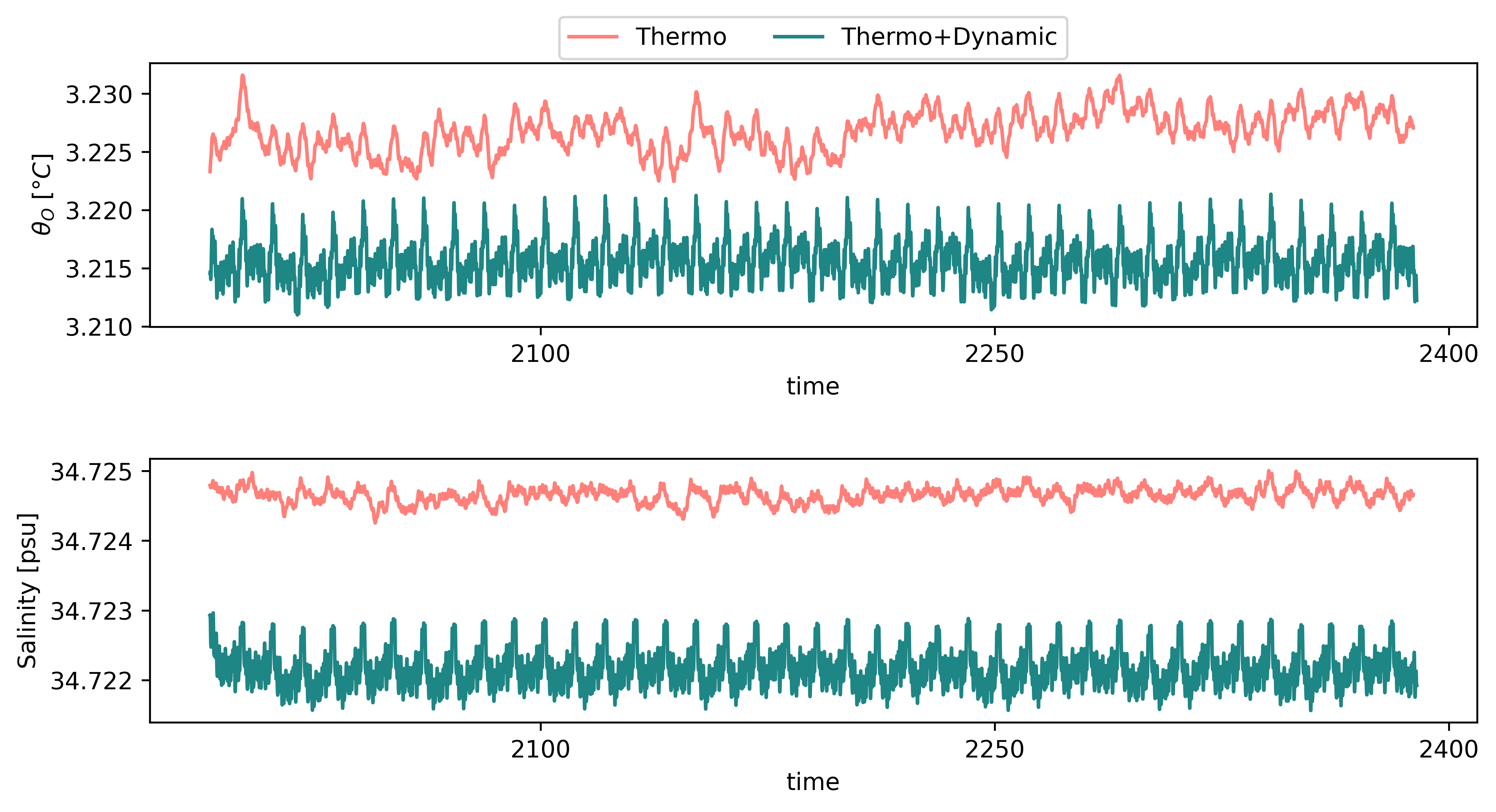}
    \caption{Spatially averaged potential temperature ($\theta_O$) and salinity ($S$) time series over a 400-year run forced with repeated atmospheric conditions taken from 1990-2000 for emulators $\mathcal{F}_{\text{thermo}}$ and $\mathcal{F}_{\text{thermo+dynamic}}$. The time series is averaged over 300-day rolling windows for visual clarity. The potential temperature trends for $\mathcal{F}_{\text{thermo}}$ and $\mathcal{F}_{\text{thermo+dynamic}}$ are $7.39 
    \times 10^{-6}$ $^{\circ} C$/year and $4.08 \times 10^{-7}$ $^{\circ} C$/year, respectively, while the Salinity trends are $1.867 \times 10^{-7}$ psu/year and $-1.397 \times 10^{-8}$ psu/year, respectively.}
    \label{fig:OM4HC}
\end{figure}

\begin{figure}
\noindent\includegraphics[width=\textwidth]{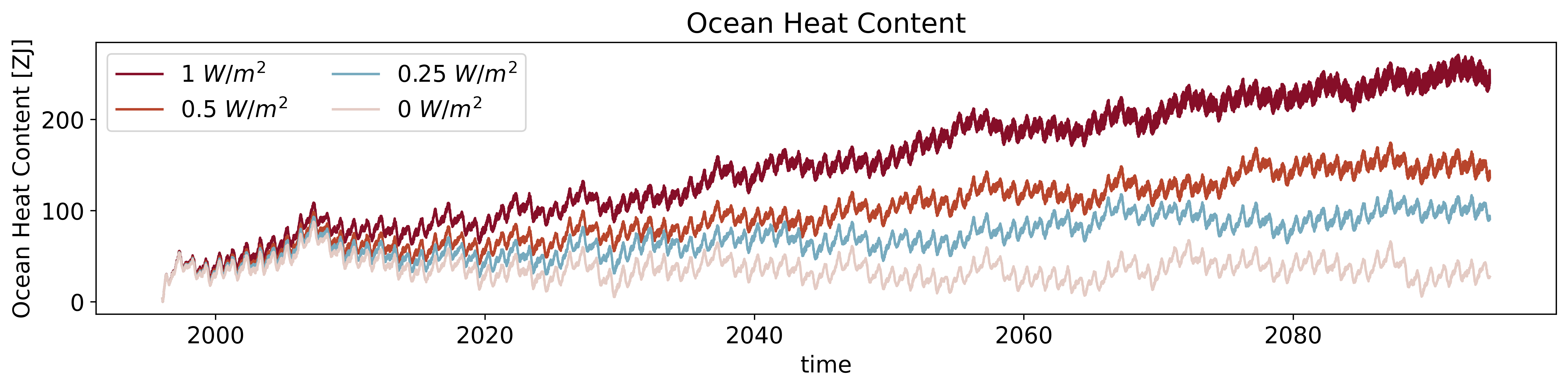}
\caption{Ocean heat content trends for 100 year runs from the $\mathcal{F}_{\text{thermo}}$ emulator. These runs are forced by increasing the global heat flux, forcing $0, 0.25,0.5,$  and  $1 W/m^2$ per year to show how the emulator responds under a range of warming conditions. }
\label{fig:fig4}
\end{figure}

\begin{figure}
\noindent\includegraphics[width=\textwidth]{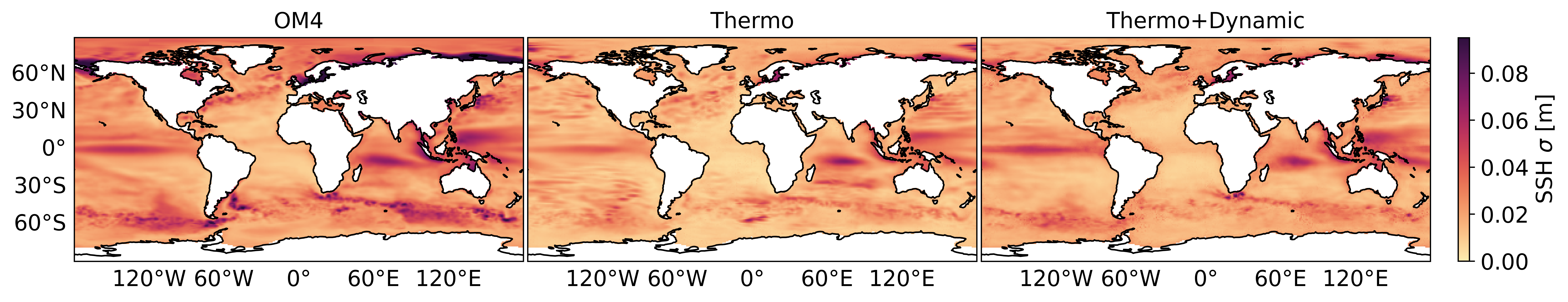}
\caption{Maps of sea surface height ($\operatorname{SSH}$), over an 8-year test set: standard deviation of anomalies, relative to climatology, for the ground truth (OM4) (left), $\mathcal{F}_{\text{thermo}}$ (center), and  $\mathcal{F}_{\text{thermo+dynamic}}$ (right).
}
\label{fig:figsshvar}
\end{figure}

\begin{figure}
\noindent\includegraphics[width=\textwidth]{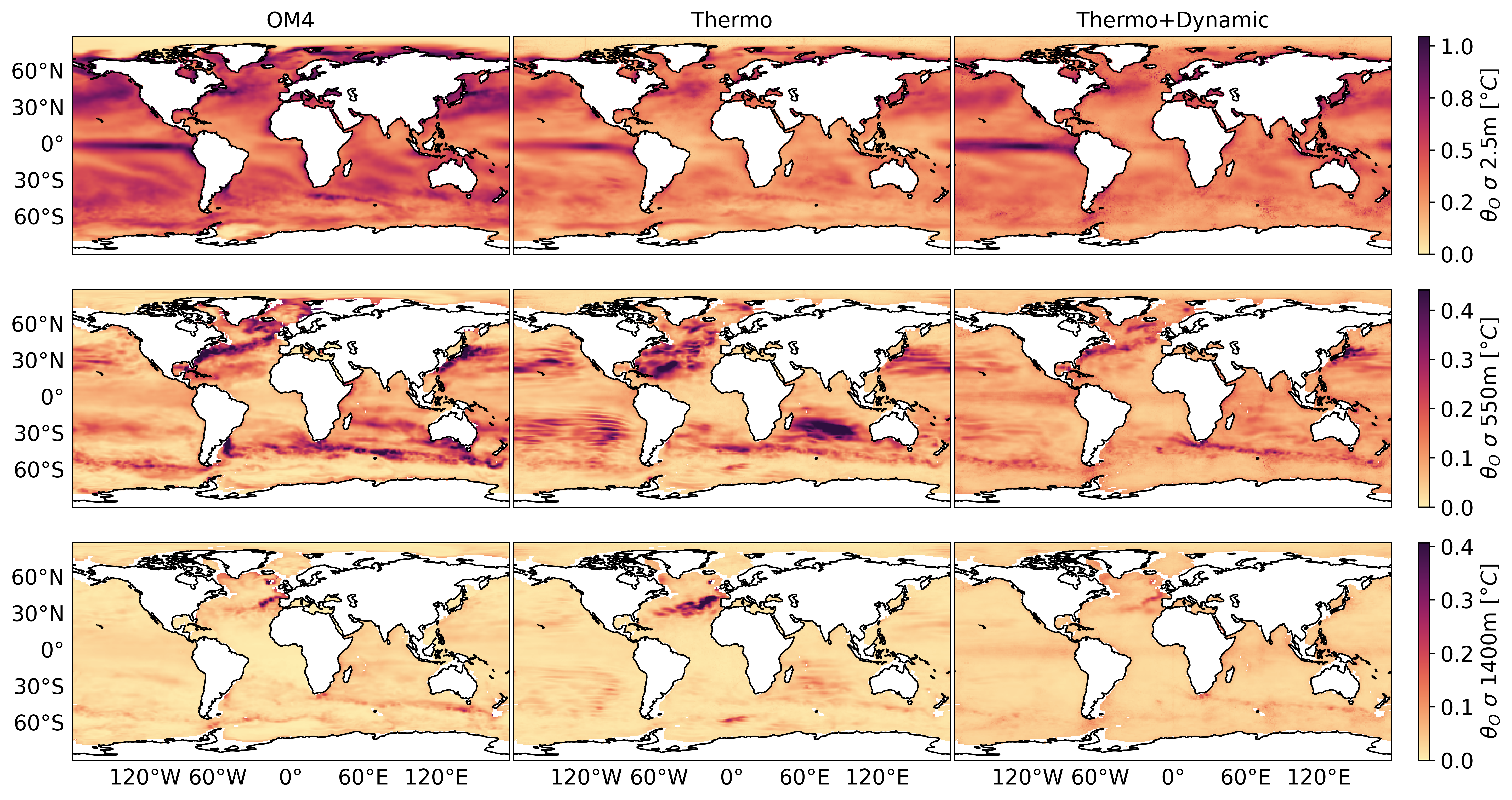}
\caption{Maps of potential temperature ($\theta_O$), over an 8-year test set at levels 2.5$\mathrm{m}$, 550$\mathrm{m}$, and 1400$\mathrm{m}$ (top to bottom): standard deviation of anomalies, relative to climatology, for the ground truth (OM4) (left), $\mathcal{F}_{\text{thermo}}$ (center), and $\mathcal{F}_{\text{thermo+dynamic}}$ (right). The emulators are exhibiting enhanced variability in the mid-latitudes and the Tropics, as expected, compared to other regions for SSH, surface temperature and surface salinity. The emulators capture the ENSO pattern of variability, Southern Ocean, and midlatitude jets. However, the amplitude of the variance is smaller than in OM4.  
At depths, the emulators have too pronounced variability in the eastern part of the North Atlantic basin, and in the Indian Ocean. 
}
\label{fig:figtempvar}
\end{figure}

\begin{figure}
\noindent\includegraphics[width=\textwidth]{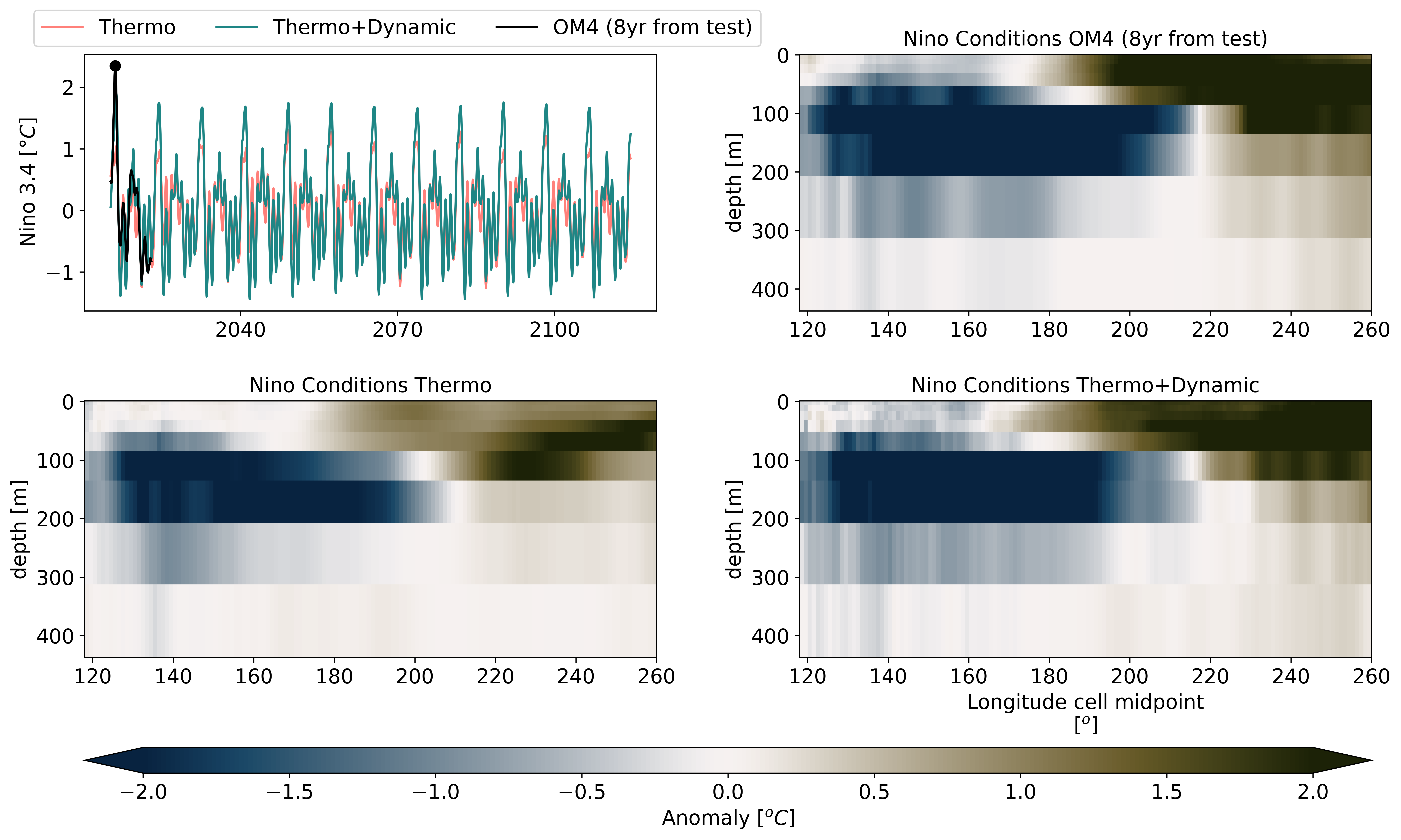}
\caption{Timeseries of Nino 3.4 index over a 100-year control run, comparing the 8-year repeat ground truth OM4 from the test set (black) with predictions from $\mathcal{F}_{\text{thermo}}$ (red) and $\mathcal{F}_{\text{thermo+dynamic}}$ (green). Here, we consider the 2.5$\mathrm{m}$ temperature anomalies. Anomalies are calculated relative to the 8-year climatology of OM4 and 100-year climatology of each emulator. Additionally, the depth structure of anomalies is shown for the peak Nina event (marked by a black dot in the timeseries).  Anomalies are averaged over rolling 150-day windows in the timeseries while the anomalies in the depth structures are averaged over a 15-day (3-snapshot) window to reduce mesoscale variability. The emulators continue to produce stable rollouts, with underestimation of magnitude similar to results obtained from test-set-only evaluations.
}
\label{fig:testlongroll}
\end{figure}

\begin{figure}
\noindent\includegraphics[width=\textwidth]{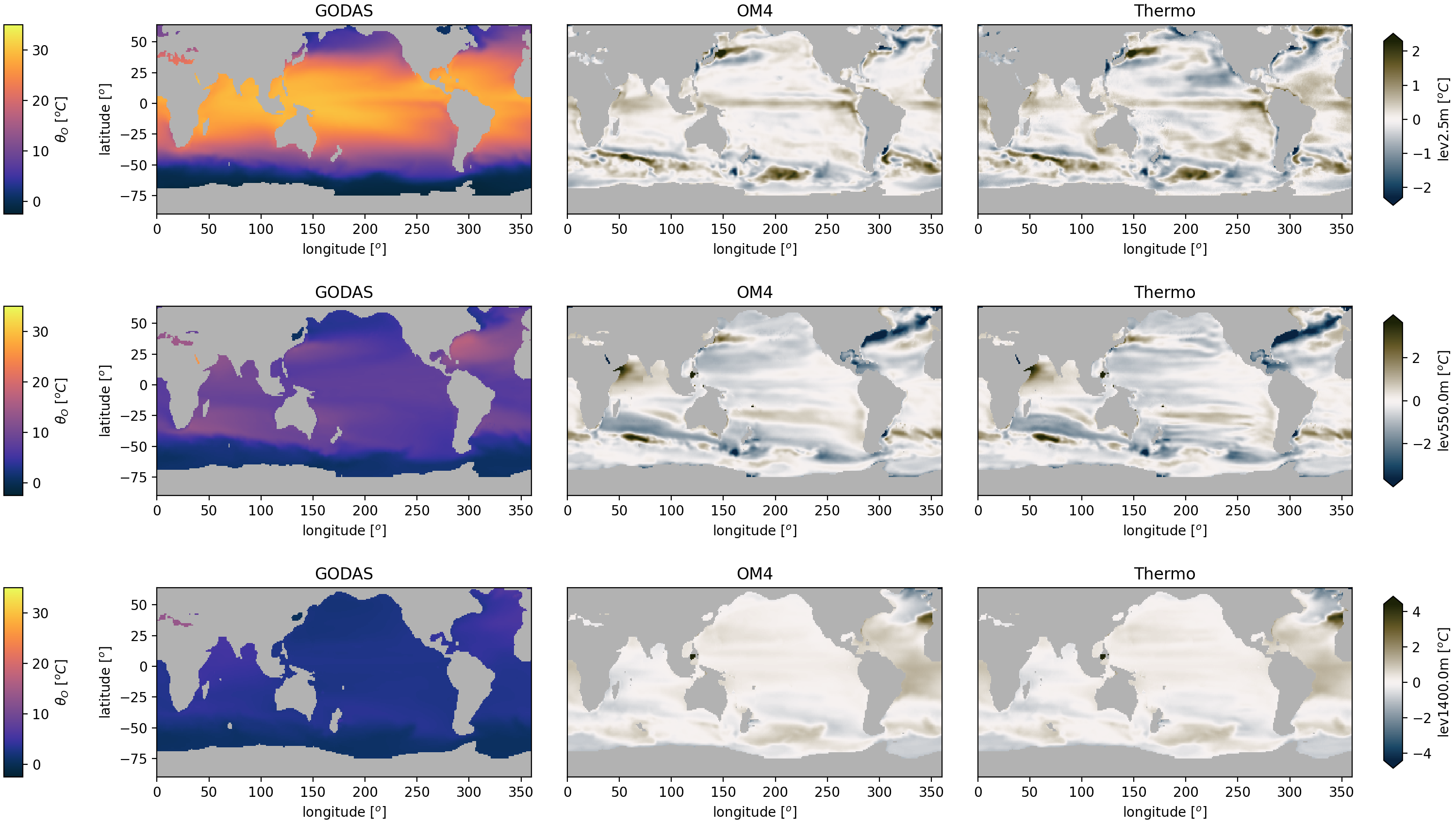}
\caption{Mean temperature bias relative to GODAS reanalysis product National Centers
for Environmental Prediction, National Weather Service, NOAA, U.S. Department of Commerce (2006) between 2014 and 2023 for OM4 and the Thermo Emulator. The GODAS product is a reanalysis product that assimilates temperature observations and synthetic salinity profiles. The GODAS data is processed onto the same horizontal and vertical grid as OM4; however, the GODAS data is only reported south $65^\circ N$. Biases from the emulator match errors produced by OM4, except in a few regions in the eastern Pacific, near the Gulf Stream, and on the boundary of the Pacific and Southern Ocean.}
\label{fig:godas_temperature_maps}
\end{figure}

\begin{figure}
\noindent\includegraphics[width=\textwidth]{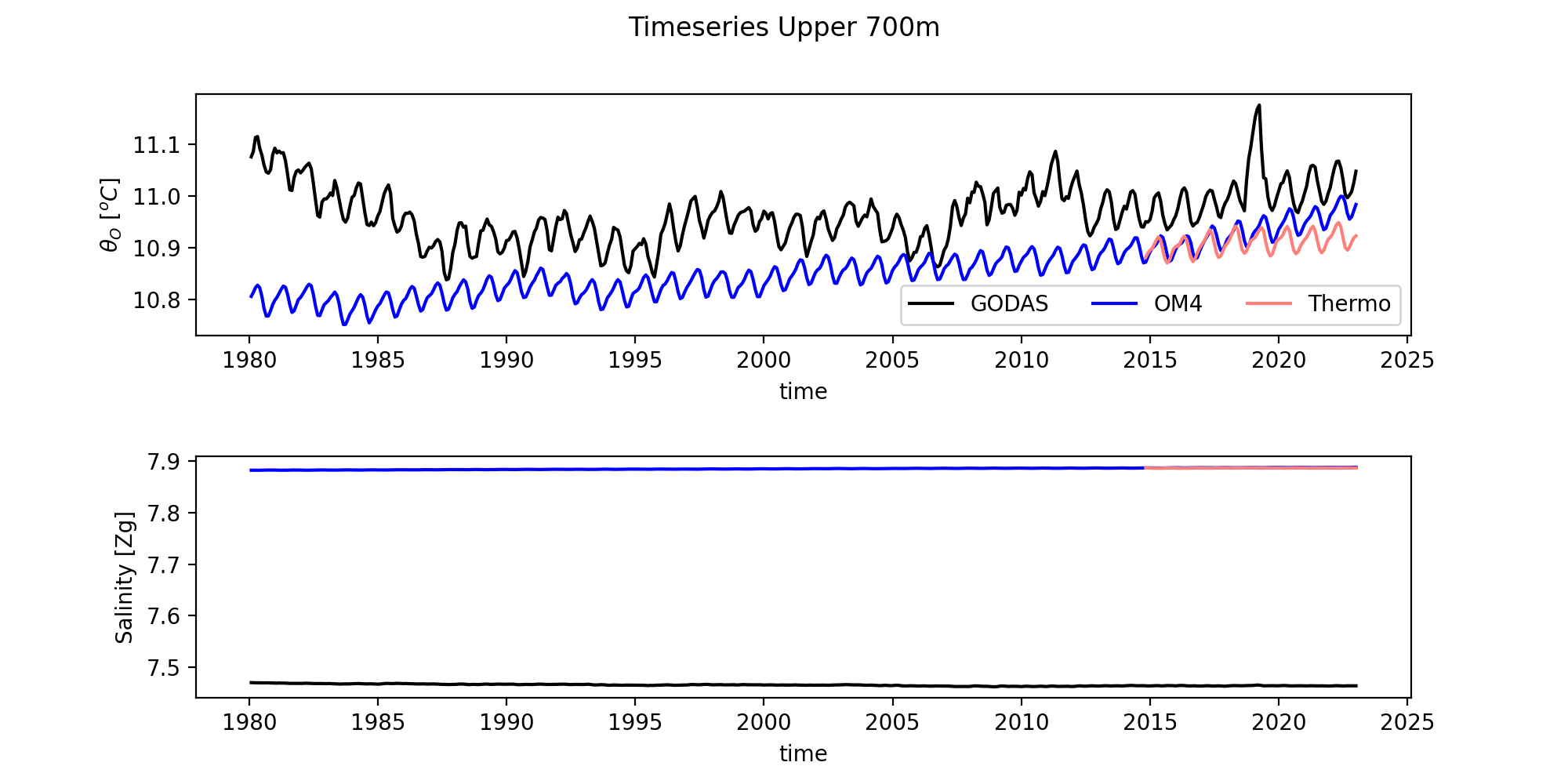}
\caption{Mean temperature and total salinity timeseries between 1980 and 2023 shown for the GODAS reanalysis (black), OM4 (blue), and the Thermo Emulator (red). The means and integrals are taken over $-85^\circ S-65^\circ N$. We see significant differences in both timeseries relative to GODAS, for the mean states, trends, and variability. As previously noted, the emulator fails to reproduce the trend over the period 2014-2023.}
\label{fig:godas_timeseries}
\end{figure}

\begin{figure}
\noindent\includegraphics[width=\textwidth]{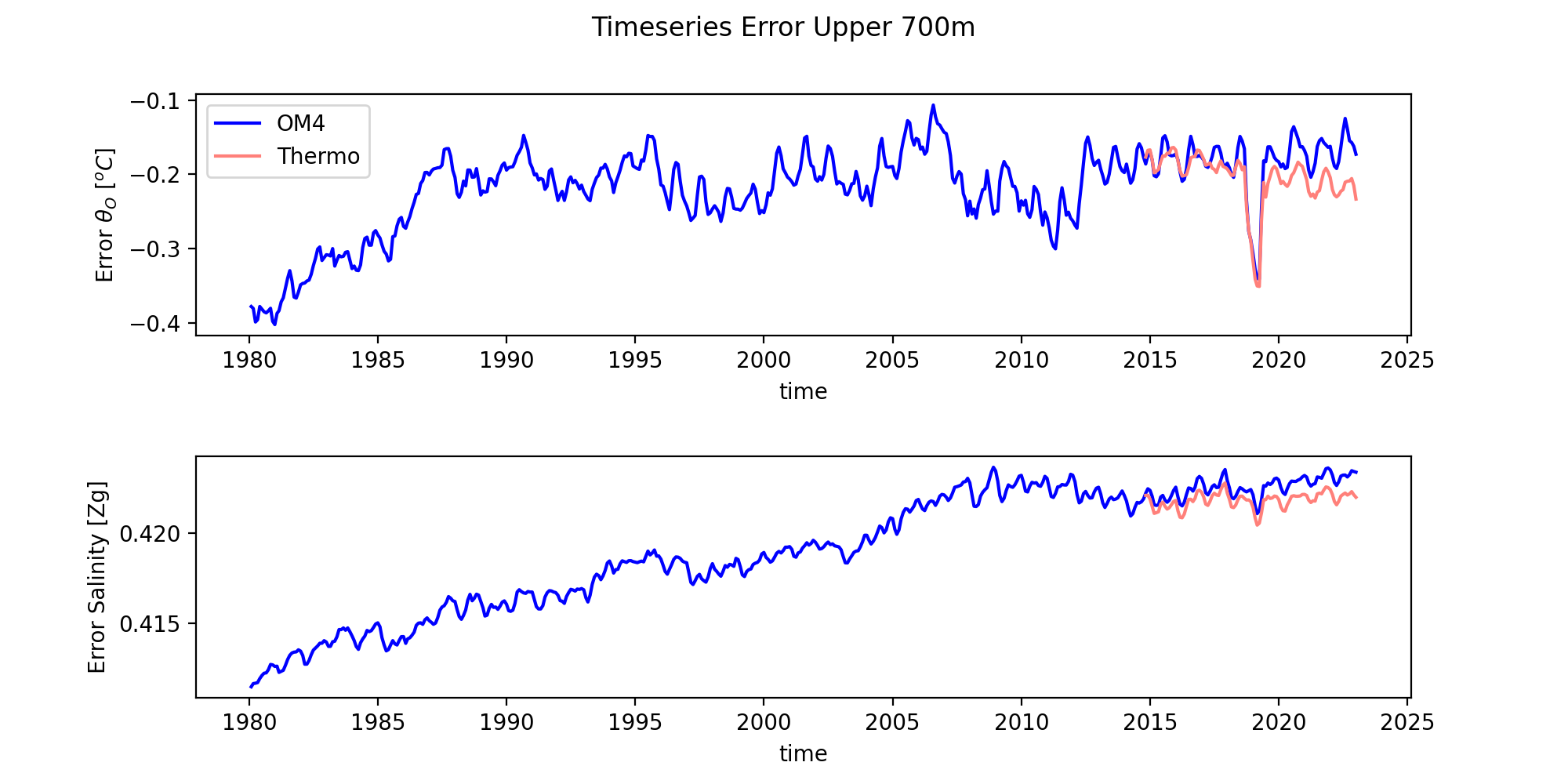}
\caption{Errors in mean temperature and total salinity timeseries between 1980 and 2023 shown for GODAS reanalysis (black), OM4 (blue) and the Thermo Emulator (red). The means and integrals are taken over $-85^\circ S-65^\circ N$. These errors relative to GODAS reflect the differences in \ref{fig:godas_timeseries}. The larger error for the emulator relative to OM4 develops from the emulator losing track of the warming trend.}
\label{fig:godas_timeseries_error}
\end{figure}

\begin{figure}
    \centering
    \includegraphics[width=\linewidth]{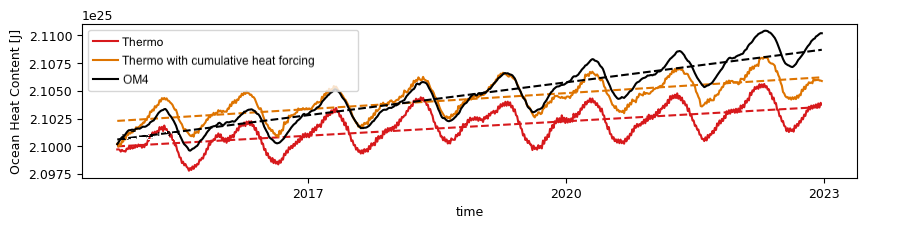}
    \caption{OHC trends for OM4 (black), $\mathcal{F}_{\text{thermo}}$ (red), and $\mathcal{F}_{\text{thermo}}$ with an additional boundary condition, given by the cumulative heat flux forcing. The trends represent the model rollout over the 8-year test set. This new model has a stronger trend compared to the original configuration, closer to the truth. However, as shown in Figure \ref{fig:cum-forcing-long}, the model produces unstable rollouts after 80 years.}
    \label{fig:cum-forcing-short}
\end{figure}

\begin{figure}
    \centering
    \includegraphics[width=\linewidth]{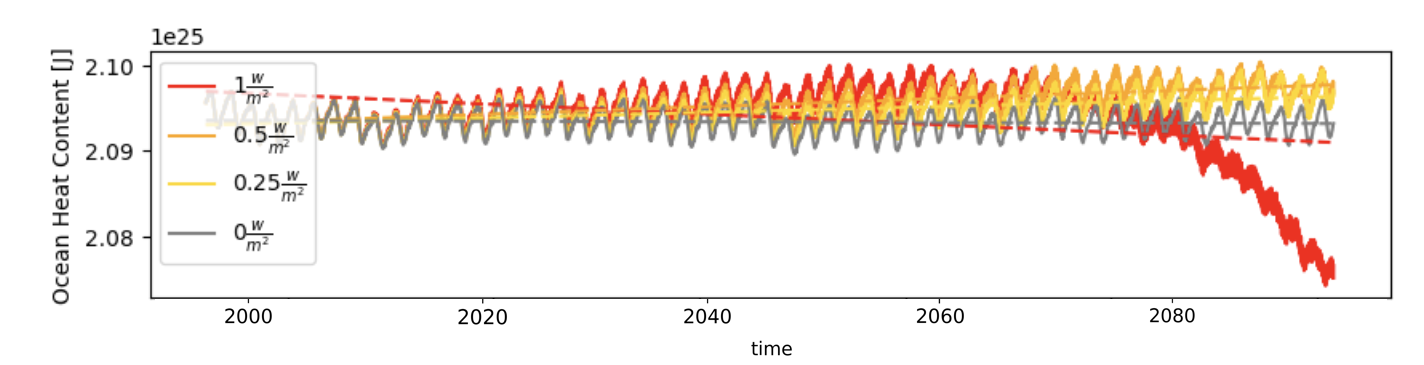}
    \caption{OHC trends (same caption as S23) for 100-year model rollout forced with increasing global heat flux, forcing 0, 0.25, 0.5, and 1$W/m^2$ per year. We show little sensitivity between the different forced runs at the start, and the stronger forcing leads to unstable results. }
    \label{fig:cum-forcing-long}
\end{figure}

\begin{figure}
    \centering
    \includegraphics[width=\linewidth]{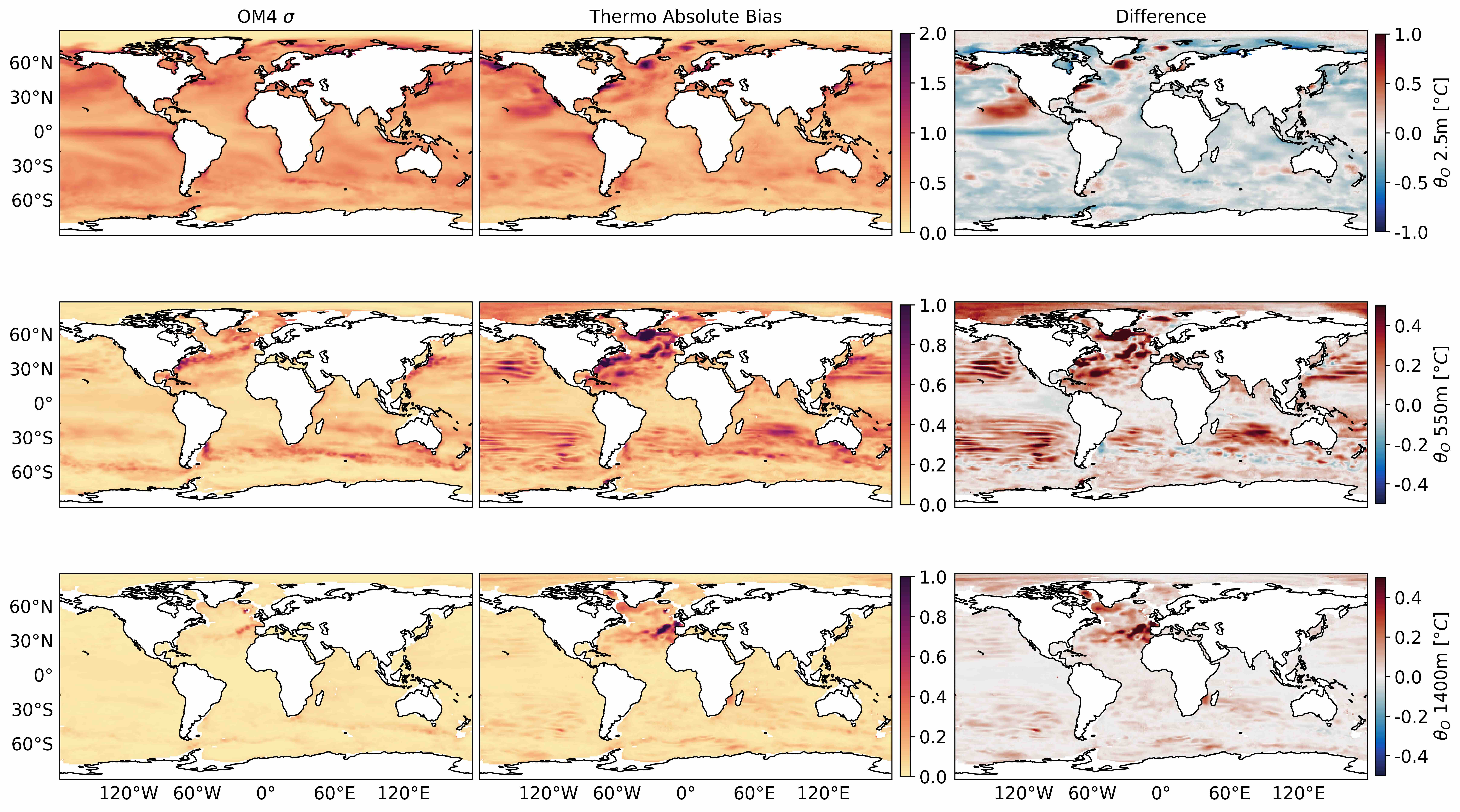}
    \caption{Maps of potential temperature ($\theta_O$), over an 8-year test set at levels 2.5$\mathrm{m}$, 550$\mathrm{m}$, and 1400$\mathrm{m}$ (top to bottom) : standard deviation of anomalies, relative to climatology for the ground truth (OM4) (left), bias of the $\mathcal{F}_{\text{thermo}}$ emulator relative to OM4 (center), and the difference between the two (i.e. center - left). At the surface, the North Atlantic and Gulf Stream regions show errors that are larger than the internal variability of the model; similarly for the region off the Californian coast. Elsewhere near the surface the errors are smaller than internal variability. In the subsurface; however, as mentioned in the main text, we see errors that are relatively large, in particular compared to the internal variability of the model. We note that the bias of the emulator for the subsurface does project onto internal variability of OM4.}
    \label{fig:temp-std-bias-diff}
\end{figure}

\begin{figure}
    \centering
    \includegraphics[width=\linewidth]{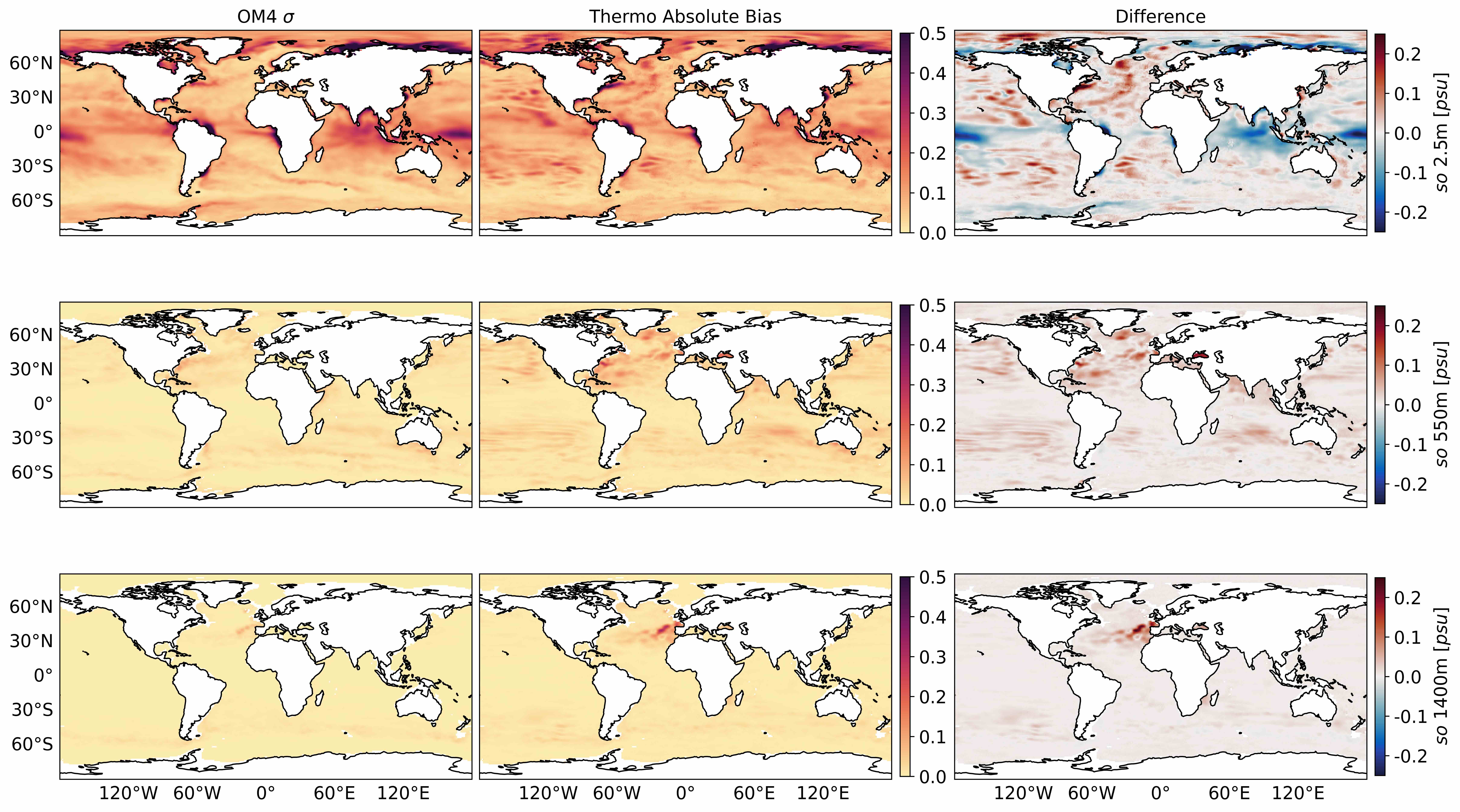}
    \caption{Maps of salinity ($so$), over an 8-year test set at levels 2.5$\mathrm{m}$, 550$\mathrm{m}$, and 1400$\mathrm{m}$ (top to bottom): standard deviation of anomalies, relative to climatology for the ground truth (OM4) (left), bias of the $\mathcal{F}_{\text{thermo}}$ emulator relative to OM4 (center), and the difference between the two (i.e. center - left), revealing the magnitude of the emulator errors relative to internal variability of the ground truth. At the surface, we can identify several regions for which the error of the emulator is larger than the internal variability, in particular in the North Atlantic. Yet, many of these regions exhibit little internal variability in the ground truth. In contrast, the errors of the emulators in the tropics are smaller than OM4 internal variability. In the subsurface, where variability is greatly diminished, the error of the emulators is larger in a few regions.
 }
    \label{fig:salinity-std-bias-diff}
\end{figure}

\begin{figure}
    \centering
    \includegraphics[width=0.5\linewidth]{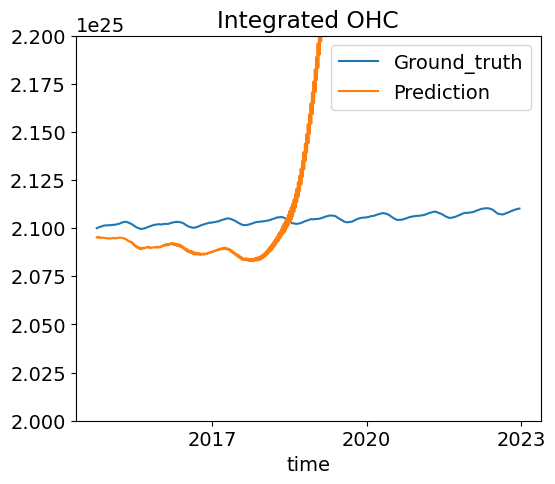}
    \caption{OHC trends for OM4 (blue) and Samudra (orange) with 2-input 1-output configuration. This plot shows a representative example of an instability during a short rollout when using one of the unstable configurations mentioned in the main text. }
    \label{fig:cum-forcing-long}
\end{figure}

\end{document}